%% file: main.tex
\documentclass[acmsmall]{acmart}

\AtBeginDocument{%
  }

\setcopyright{cc}
\setcctype{by}
\acmDOI{10.1145/3797106}
\acmYear{2026}
\acmJournal{PACMSE}
\acmVolume{3}
\acmNumber{FSE}
\acmArticle{FSE078}
\acmMonth{7}
\acmSubmissionID{fse26maina-p2570-p}
\received{2025-09-12}
\received[accepted]{2025-12-22}

\usepackage{multirow}
\usepackage[most]{tcolorbox}
\usepackage{pifont}
\usepackage{graphicx}

\usepackage{xspace}
\usepackage{enumitem}
\usepackage{colortbl}

\usepackage{hyperref}
\usepackage{algorithm}
\usepackage{algpseudocode}
\usepackage{array} 
\usepackage{multirow}
\usepackage{xcolor}
\usepackage{threeparttable}
\usepackage[most]{tcolorbox}
\usepackage{listings}
\usepackage{lipsum}
\tcbuselibrary{listingsutf8}
\usepackage{nameref}
\usepackage[utf8]{inputenc}

\usepackage{caption} 

\usepackage{booktabs}      
\usepackage{graphicx}      

\definecolor{heatmapgreen}{RGB}{173, 216, 230}
\newcommand{\colorcell}[1]{%
  \ifdim#1pt>40pt
    \cellcolor{heatmapgreen!#1}\color{black}#1%
  \else
    \cellcolor{heatmapgreen!#1}\color{black}#1%
  \fi
}

\definecolor{heatmapblue}{RGB}{173, 216, 230}
\newcommand{\rolorcell}[1]{%
  \ifdim#1pt>1.6pt
    \cellcolor{heatmapblue!#1}\color{black}#1%
  \else
    \cellcolor{heatmapblue!#1}\color{black}#1%
  \fi
}

\tcbuselibrary{listings,breakable}
\newcounter{myboxctr}
\newtcolorbox{mybox}[3][]{
  floatplacement={#3},
  float,
  enhanced,
  colback=white, colframe=black!50,
  left=8pt,
  top=2pt,
  right=7pt,
  bottom=1pt,
  after skip=2pt,
  fonttitle=\small\bfseries,
  fontupper=\scriptsize,
  breakable,
  before title={\refstepcounter{myboxctr}\label{#1}},
  title=Box~\themyboxctr: #2,
}

\usepackage[framemethod=tikz]{mdframed}
\mdfdefinestyle{customstyle}{
  linecolor=gray!100,
  linewidth=3pt,
  innerleftmargin=3pt,
  topline=false,
  rightline=false,
  bottomline=false,
  leftline=true,
  innerrightmargin=3pt,
  innertopmargin=3pt,
  innerbottommargin=3pt,
  backgroundcolor=gray!15
}
\newenvironment{custommdframed}
  {\begin{mdframed}[style=customstyle]}
  {\end{mdframed}}

\newcommand{\cmark}{\textcolor{green!60!black}{\ding{51}}} 
\newcommand{\xmark}{\textcolor{red}{\ding{55}}} 
\definecolor{DeepGreen}{RGB}{0,128,0}
\definecolor{DeepRed}{RGB}{178,34,34}
\definecolor{lightgray}{gray}{0.92}

\definecolor{codegreen}{rgb}{0,0.6,0}
\definecolor{codegray}{rgb}{0.5,0.5,0.5}
\definecolor{codepurple}{rgb}{0.58,0,0.82}
\definecolor{backcolour}{rgb}{0.95,0.95,0.96}
\definecolor{framecolor}{rgb}{0.8,0.8,0.8}

\usepackage{xcolor} 
\usepackage[most]{tcolorbox} 

\definecolor{lightyellow}{rgb}{1.0, 1.0, 0.9} 
\tcbset{
    colback=lightyellow, 
    colframe=lightyellow, 
    arc=0mm, 
    auto outer arc=true, 
    boxsep=0pt, 
    left=5pt, right=5pt, top=5pt, bottom=5pt, 
    nobeforeafter 
}

\newcommand{\eg}{{\emph{e.g.,}}\xspace}
\newcommand{\ie}{{\emph{i.e.,}}\xspace}

\begin{document}

\title[A Repo-Level Code Generation Benchmark Aligned with Real-World Software Development Practices]{RealBench: A Repo-Level Code Generation Benchmark Aligned with Real-World Software Development Practices}

\author[Jia. Li, Hongyi. Deng et al.]{Jia Li}
\orcid{0000-0002-9411-971X}
\affiliation{%
  \institution{ Wuhan University}
  \city{Wuhan}
  \country{China}
}
\email{jia.li@whu.edu.cn}

\author{Hongyi Deng}
\orcid{0009-0007-6658-2823}
\affiliation{%
  \institution{Peking University}
  \city{Beijing}
  \country{China}
}
\email{2401210232@stu.pku.edu.cn}

\author{Yiran Zhang}
\orcid{0000-0002-9366-6076}
\authornote{Corresponding authors.}
\affiliation{%
  \institution{Nanyang Technological University}
  \city{Singapore}
  \country{Singapore}
}
\email{yiran002@e.ntu.edu.sg}

\author{Kechi Zhang}
\orcid{0000-0002-3290-0244}
\affiliation{%
  \department{Key Lab of High Confidence Software Technology (PKU), Ministry of Education, School of Computer Science,}
  \institution{Peking University}
  \city{Beijing}
  \country{China}
}
\email{zhangkechi@pku.edu.cn}

\author{Tianqi Shao}
\orcid{0009-0008-1503-5077}
\affiliation{%
  \institution{Peking University}
  \city{Beijing}
  \country{China}
}
\email{shaotianqi@stu.pku.edu.cn}

\author{Tiankuo Zhao}
\orcid{0009-0004-1869-7106}
\affiliation{%
  \institution{ Wuhan University}
  \city{Wuhan}
  \country{China}
}
\email{tiankuz@whu.edu.cn}

\author{Weinan Wang}
\orcid{0009-0002-5762-6277}
\affiliation{%
  \institution{Peking University}
  \city{Beijing}
  \country{China}
}
\email{wangwn@pku.edu.cn}

\author{Zhi Jin}
\orcid{0000-0003-1087-226X}
\authornotemark[1]
\affiliation{%
  \institution{Wuhan University}
  \city{Wuhan}
  \country{China}
}
\email{zhijin@whu.edu.cn}

\author{Ge Li}
\orcid{0000-0002-5828-0186}
\affiliation{%
  \department{Key Lab of High Confidence Software Technology (PKU), Ministry of Education, School of Computer Science,}
  \institution{Peking University}
  \city{Beijing}
  \country{China}
}
\email{lige@pku.edu.cn}

\author{Yang Liu}
\orcid{0000-0001-7300-9215}
\affiliation{%
  \institution{Nanyang Technological University}
  \city{Singapore}
  \country{Singapore}
}
\email{yangliu@ntu.edu.sg}

\author{Yingtao Fang}
\orcid{0009-0000-8253-226X}
\affiliation{%
  \institution{Wuhan University}
  \city{Wuhan}
  \country{China}
}
\email{19855359197@163.com}

\author{Yihong Dong}
\orcid{0000-0001-6228-4019}
\affiliation{%
\department{Key Lab of High Confidence Software Technology (PKU), Ministry of Education, School of Computer Science,}
  \institution{Peking University}
  \city{Beijing}
  \country{China}
}
\email{dongyh@stu.pku.edu.cn}

\renewcommand{\shortauthors}{J. Li, H. Deng, Y. Zhang, K. Zhang, T. Shao, T. Zhao, W. Wang, Z. Jin, G. Li, Y. Liu, Y. Fang, Y. Dong.}








\begin{abstract}
 
Writing code requires significant time and effort in software development. To automate this process, researchers have made substantial progress using Large Language Models (LLMs) for code generation. Many benchmarks like HumanEval and EvoCodeBench have been created to evaluate LLMs by requiring them to generate code from natural language requirements. However, in enterprise applications and team development, developers typically write code based on structured designs or specifications rather than raw natural language descriptions. This gap between existing benchmarks and real industry development practices means that current benchmark scores may not accurately reflect how much code generation can help automate software development tasks.

To address this gap, we propose RealBench, a repository-level code generation benchmark aligned with real-world industry software development practices. Each example includes both natural language requirements and UML diagrams as system design, matching how developers typically receive specifications.  Based on the constructed benchmarks, we conduct a systematic evaluation of advanced LLMs' code generation capabilities when provided with structured system designs. Specifically, we design three generation strategies to evaluate advanced LLMs on RealBench and propose two evaluation granularities with five metrics. The experimental results reveal key insights in current LLMs' capabilities for repo-level code generation aligned with real-world software development practices. First, we notice that regarding repo-level code generation, LLMs show much worse performance and there are significant performance gaps among LLMs. Second, LLMs are good at finding and creating modules defined in UML diagrams, but the quality of generated modules is often poor due to grammar and logic errors. Third, generating the entire repository at once is the best generation strategy on smaller repositories, while generating a complex repository with the module-by-module strategy works better compared to other strategies. Fourth, the detailed system design is very important for repository-level code generation tasks through conducting ablation studies on system designs. Lastly, we discuss the frequent error types in generated repositories to provide insights for optimizing repo-level code generation.

\end{abstract}





\maketitle
\input{sec1-introduction}

\input{sec2-benchmark}

\input{sec3-Generation}

\input{sec4-experiments}

\input{sec5-analysis}
\input{sec6-relatedwork}

\bibliographystyle{ACM-Reference-Format}
\bibliography{main}


\end{document}

%% file: sec1-introduction.tex
\section{Introduction}\label{sec:intro}

Writing code takes significant time and effort in software development. To make this process easier, researchers have spent significant efforts on automated code generation. Recent progress in Large Language Models (LLMs) such as GPT-4o \cite{GPT-4o}, DeepSeek \cite{liu2024deepseek}, and Qwen \cite{bai2023qwen} has shown promising results for generating code automatically \cite{wei2024selfcodealign, li2024deveval}.

To better understand how well LLMs generate code, researchers have created many benchmarks like HumanEval \cite{chen2021evaluating} and EvoCodeBench \cite{li2024evocodebench}. These benchmarks mainly focus on two types of code generation scenarios. First, many benchmarks \cite{chen2021evaluating, iyer2018mapping} focus on the simple code generation scenario, \ie function-level or statement-level code generation, which is also the most common one. They mainly require LLMs to generate a single code unit (\eg a function or a statement) based on a given natural language requirement. Such evaluation aims to generate code with short length. For example, each target code in the widely-used benchmark HumanEval only contains 11.5 lines and 24.4 tokens on average, where GPT-4o and DeepSeek-V3 can achieve 92.7\% and 91.5\% test pass rate performance on it. Second, with the ability and context window length of LLMs improving, researchers introduce several repo-level code generation benchmarks \cite{li2024evocodebench, li2024deveval, cao2024javabench}. It asks LLMs to generate a function/class (\eg EvoCodeBench \cite{li2024evocodebench} and ClassEval \cite{du2023classeval}) based on a natural language requirement and a repository, or even an entire repository (\eg CODES \cite{zan2024codes} and JavaBench \cite{cao2024javabench}) based on requirements. For the scenario of generating a function/class, LLMs understand the dependency relations of code snippets in the current repository and predict code.

However, \textbf{these benchmarks differ from common development practices in real-world industry settings.}
In waterfall development, developers write code from a structured design that has been translated from stakeholder requirements, rather than from the raw requirements themselves.
This creates a significant gap between existing benchmarks and real development scenarios. As a result, benchmark scores may not show how much code generation can truly help automate real software development.

To mitigate this gap, we propose RealBench, a repository-level code generation benchmark aligned with real-world software development practices. 
Table \ref{tab:benchmark_comparison} shows the comparison between our benchmark and mainstream code generation benchmarks.
Our benchmark offers the following key features:

\begin{itemize}[leftmargin=*]
    \item  \textbf{Align with Industry Software Development Practices.} Each example provides both natural language requirements and UML (Unified Modeling Language) diagrams as system design, instead of only requirements, matching how developers receive specifications in practice. 
    \item \textbf{Real-world Repository Across Diverse Domains.} The benchmark is collected from real-world code repositories. It consists of 61 repositories from 20 popular programming domains (\eg Internet, Security, and Database), instead of being composed of randomly selected a few repositories like some existing benchmarks. 
    \item  \textbf{Different Difficulty Levels.} RealBench contains four levels based on the line of code (LOC) in repositories, including level-1 (0$\sim$500 LOC), level-2 (500$\sim$1000 LOC), level-3 (1000$\sim$2000 LOC), and level-4 ($\geq$2000 LOC). 
    \item  \textbf{Reduced Data Contamination.}  We collect up-to-date code repositories that are created after 2024-12 from GitHub\footnote{https://github.com/}, which are later than most prevailing LLMs' cut-off dates thus reducing the risk of data contamination.
    \item \textbf{Comprehensive Test Suite.} RealBench provides comprehensive human-verified test suites with an average of 50 test cases per repository and 79.76\% line coverage.
\end{itemize}

\begin{table}[htbp]
  \centering
  \caption{Comparison of Existing Code Generation Benchmarks. \#Req and \#Diag mean requirements and UML diagrams, respectively.  \#Real-R represents whether benchmarks are collected from real-world repositories. \#Func means the average number of functions in each example. \#Repo means whether benchmarks have repo-level evaluation metrics.}
  \resizebox{\linewidth}{!}{
    \begin{tabular}{l|c|cc|cccc|cc|c|ccc}
    \toprule
    \multicolumn{1}{l|}{\multirow{2}[4]{*}{Benchmark}} & \multicolumn{1}{c|}{Task} & \multicolumn{2}{c|}{Input Information} & \multicolumn{4}{c|}{Different Difficult Levels} & \multicolumn{2}{c|}{Source} & \multicolumn{1}{c|}{Data Leaking} & \multicolumn{3}{c}{Evaluation} \\
     & \multicolumn{1}{c|}{Granularity}    & \multicolumn{1}{c}{\#Req} & \multicolumn{1}{c|}{\#Diag} & \multicolumn{1}{c}{\#Func} & \multicolumn{1}{c}{Class} & \multicolumn{1}{c}{File} & \multicolumn{1}{c|}{\#LOC} & \multicolumn{1}{c}{\#Real-R}  & \multicolumn{1}{c|}{\#Num} & \multicolumn{1}{c|}{Data Time} & \multicolumn{1}{c}{\#Repo} & \multicolumn{1}{c}{Class/\#Func}  & \multicolumn{1}{c}{Tests} \\
    \midrule
      Concode~\cite{iyer2018mapping} &  Function & \cmark  &  \xmark  & 2,000 &0  & 0 &  --   & \cmark   & --  &  2018 &  \xmark  & \xmark  &-- \\
      CoNaLA~\cite{yin2018learning} &  Statement    &   \xmark    &   \xmark & 500 &0  & 0 & 1  & \cmark   & --  & 2018    &  \xmark & \xmark   & -- \\
     APPS~\cite{hendrycks2021measuring} & Function  &\cmark  &  \xmark   & 5,000 & 0 & 0 &21.4  & \xmark   & -- & 2021 &  \xmark & \cmark & 13.2  \\
    HumanEval~\cite{chen2021evaluating} & Function   & \cmark  &  \xmark  & 164  & 0  & 0  & 11.5 &\xmark  & -- &2021   & \xmark & \cmark & 7.7 \\
    MBPP~\cite{austin2021program} & Function & \cmark  &  \xmark & 974  & 0  & 0  & 6.8 & \xmark  & -- & 2021 & \xmark & \cmark & 3.0 \\
    math-qa~\cite{austin2021program}   & Statement  &  \xmark     & \xmark  & 2,985 & 0 &0 & 7.6 & \cmark  & -- & 2021   & \xmark  & \xmark   & -- \\
    HumanEval-X~\cite{zheng2023codegeex} & Function  & \cmark  & \xmark & 164  & 0  & 0 & 11.5 & \xmark   & --  & 2022    & \xmark  & \cmark & 7.7 \\
    MBXP~\cite{athiwaratkun2022multi} & Function &\cmark  & \xmark &  974 & 0  & 0 & 6.8 & \xmark  & --  & 2022  & \xmark  &  \cmark  & 3.0  \\
    multi-math-qa~\cite{athiwaratkun2022multi} & Statement   & \xmark  & \xmark  & 2,985 & 0  & 0 & 7.6  & \xmark   & --    & 2022 & \xmark  &  \xmark & --  \\
    CodeContests~\cite{li2022competition} & Function  & \cmark  & \xmark & 165 & 0  & 0 & 59.8 & \cmark  & --  &  2022  & \xmark   & \cmark & 1.2 \\
    PandasEval~\cite{zan2022cert} &  Function & \cmark &\xmark  & 101 &  0  &0 & 1.3 & \xmark  & -- & 2022 & \xmark   &  \cmark  & 6.5 \\
    NumpyEval~\cite{zan2022cert}  & Function  & \cmark &\xmark  &  101  & 0  & 0 & 1.1 & \xmark  & -- &2022  & \xmark  &  \cmark  & 3.5  \\
    DS-1000~\cite{lai2023ds}  & Statement  & \xmark  &\xmark  & 1,000 & 0 & 0 & 3.8 & \xmark  & -- & 2022 & \xmark & \cmark & 1.6    \\
    MTBP~\cite{nijkamp2022codegen}  & Function  & \cmark  &\xmark  & 115  & 0 & 0  & -- &   \xmark  & --   & 2022  & \xmark  &\xmark   & -- \\
   ODEX~\cite{wang2022execution}  & Function & \cmark  &\xmark  & 945  &0 & 0  & 1.9 &\xmark  & -- & 2022  & \xmark  & \cmark & 1.8  \\
   BIG-Bench~\cite{srivastava2023beyond} &Function  & \cmark  &\xmark   & 32  & 0  & 0 &  --  & \cmark & --  & 2023  & \xmark & \cmark & 4.69    \\
   CoderEval~\cite{yu2024codereval}  & Function  & \cmark  &\xmark  & 460  & 0  & 0 &  $\le$ 32 &\xmark   & -- & 2023  &\xmark   & \xmark & --    \\
    CrossCodeEval~\cite{ding2023crosscodeeval}  & Statement  &  \xmark  &\xmark &--  & 3,534 &  0 & 96.2  & \xmark   & -- & 2023  & \xmark   &  \xmark  & --  \\
   ClassEval~\cite{du2023classeval}  & Class  &  \cmark & \xmark &  412  & 100  &0  & 45.7   & \cmark  & --  & 2023 & \xmark  & \cmark & 33.1 \\
   RepoEval~\cite{zhang2023repocoder}  & Repository  & \cmark & \xmark  &1,973  & --  &--  &  $\le$30  & \cmark  & --  & 2023  &  \xmark  & \xmark & --    \\
   DevEval~\cite{li2024deveval} & Repository  & \cmark & \xmark  & 1,874  &   --     &     --   & 392.7  & \cmark   & 117 & 2024   & \xmark & \cmark  & -- \\
   EvoCodeBench~\cite{li2024evocodebench} & Repository & \cmark & \xmark  &  --     &   275  &  -- &  --  &  \cmark  & 25  &  2024-03  &  \xmark &  \cmark    &  --   \\
   CodeAgentBench~\cite{zhang2024codeagent}  &  Repository & \cmark & \xmark  &       101       &--&    --    &    --    &     \cmark   &    6   &  2024    &   \xmark   &   \cmark  &  --    \\
   OOPEval~\cite{wang2024oop} & Repository  & \cmark & \xmark & 0  & 431  &   --      &    --     &     \xmark  &  -- &  2024 & \xmark  & \cmark  & 2.48 \\
    CODES~\cite{zan2024codes}  & Repository  & \cmark & \xmark  & --  & --    &  8.43     &    --    & \cmark  & 19    &   2023-08    &  \xmark     &    \cmark    & \xmark   \\
   JavaBench~\cite{cao2024javabench}  & Repository & \cmark & \xmark  & 389   & 106 &  -- & -- & \xmark  & 4 & 2024   & \xmark  & \cmark  & 3.73 \\
   \rowcolor{cyan!20}
   RealBench (Ours)  & Repository  & \cmark  & \cmark  & 2,484 &  544  &  538  & 1,201  & \cmark   & 61  & 2024.12-2025.05 & \cmark  & \cmark   &  50.05  \\
    \bottomrule
    \end{tabular}%
  \label{tab:benchmark_comparison}%
  }
\end{table}%

Based on the constructed benchmark, we did a comprehensive study on the code generation capability of existing LLMs. Specifically, we evaluate the generated repositories at \textbf{two evaluation granularities}, including repository-level (\ie Requirement@k and Architecture@k) and class-level (\ie Completion rate, Execution pass rate, and Test pass rate) using \textbf{five metrics} evaluations.
We \textbf{evaluate 6 popular LLMs}, which contain 3 proprietary models (\ie GPT-4o \cite{GPT-4o}, Claude-Sonnet-4 \cite{Claude-Sonnet-4}, and Gemini-2.5-Flash \cite{Gemini-2.5-flash}) and 3 open source models (\ie Deepseek-V3 \cite{guo2024deepseek}, Qwen3-235B-A22B \cite{Qwen3-235B-A22B}, and Qwen2.5-Coder-7B-Instruct \cite{Qwen2.5-Coder-7B}) on RealBench. We design \textbf{three generation strategies} (\ie holistic, incremental, and retrieval-augmented generation). We conduct \textbf{ablation studies on UML diagrams} (\ie package diagram and class diagram) as the system design.

The experimental results reveal key limitations in current LLMs' capabilities for repo-level code generation aligned with real-world software development practices. The best average test pass rate score is only 19.39\% among all studied LLMs. Moreover, LLMs perform much worse as repository size grows: test pass rate scores drop from over 40\% on repositories with less than 500 lines of code to under 15\% on repositories with more than 2000 lines of code. Our further analysis shows that for smaller repositories with less than 1000 lines of code, using a holistic generation strategy works best. For larger repositories with more than 1000 lines of code, the incremental strategy works better. Our analysis also confirms that detailed design is very important for repository-level code generation. By analyzing the detailed limitations of current LLMs, we aim to provide valuable insights for optimizing repo-level code generation and driving advancements in software engineering.

Our contributions are summarized as follows.

\begin{itemize}[leftmargin=*]
    \item We propose a repo-level code generation benchmark, RealBench, aligned with real-world software development practices. It provides both natural language requirements and UML diagrams as system designs, rather than requiring LLMs to directly generate code from raw requirements.
    \item  We design a systematic evaluation design for comprehensively assessing LLMs on repo-level code generation tasks, being a pioneering evaluation practice. The assessment system contains two evaluation granularities with five metrics. 
     \item We conduct extensive experiments to observe LLMs' performance in generating repositories aligned with real-world industry software development practice. By analyzing results, we provide some valuable insights for optimizing repo-level code generation and LLMs.
\end{itemize}

%% file: sec2-benchmark.tex
\section{RealBench Benchmark}

In this section, we introduce RealBench. We will first introduce the data format, then the construction procedure, and finally the characteristics of the constructed benchmark.

\begin{figure}[tb]
    \centering
    \includegraphics[width=0.95\textwidth]{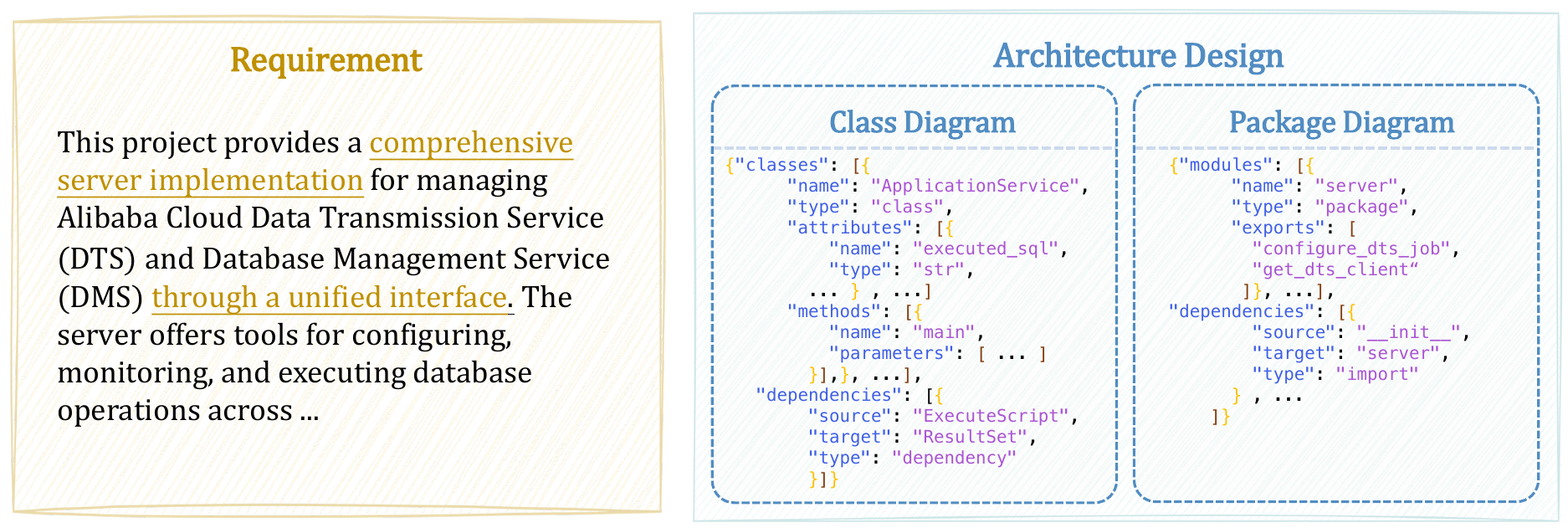}
    \caption{An Example for a Code Generation Task in RealBench.}
    \label{fig_benchmark_case}
\end{figure}

\subsection{Benchmark Format}

The specification of Realbench's content is shown in Table~\ref{tab:content_spec}. A repository comprises the natural language requirements of the whole repository and its \textit{\textbf{system design}}. The requirements specify the purpose of the repository.
For the system design, RealBench provides a two-level standard UML diagram following UML specification~\cite{OMG2017UML} for each repository: the package diagram and class diagram, where the package diagram serves as a high-level overview of the system, and the class diagram shows the detailed design. We provide an example of our benchmark in Figure~\ref{fig_benchmark_case}.

\begin{table}[t]
\centering
\scriptsize
\caption{Content Specification for RealBench.}
\label{tab:content_spec}

\begin{tabular}{ll|ll|l}
\toprule
\multicolumn{2}{c|}{\textbf{Category}}                                                                                                                                     & \multicolumn{2}{l|}{\textbf{Elements}}                      & \textbf{Definition}                                                        \\ \midrule
\multicolumn{2}{l|}{ \textit{Repository Requirements}}                                                                                                                                  & \multicolumn{2}{l|}{Repository Description}                    & Natural language specification of repository's purpose, scope, and objectives   \\ \midrule
\multicolumn{1}{l|}{\multirow{9}{*}{\begin{tabular}[c]{@{}l@{}} \textit{System}\\ \textit{Design}\end{tabular}}} & \multirow{3}{*}{\begin{tabular}[c]{@{}l@{}}Package\\ Diagram\end{tabular}} & \multicolumn{2}{l|}{Package Names}                          & Module identifiers representing logical system organization                \\ 
\multicolumn{1}{l|}{}                                                                         &                                                                            & \multicolumn{2}{l|}{Dependencies}                           & Inter-package import relationships showing architectural structure         \\ 
\multicolumn{1}{l|}{}                                                                         &                                                                            & \multicolumn{2}{l|}{Public Exports}                         & Interface elements (functions/classes) made available by each package      \\ \cmidrule{2-5} 
\multicolumn{1}{l|}{}                                                                         & \multirow{6}{*}{\begin{tabular}[c]{@{}l@{}}Class\\ Diagram\end{tabular}}   & \multicolumn{2}{l|}{Class Names}                            & Complete class definitions including utility classes from module functions \\
\multicolumn{1}{l|}{}                                                                         &                                                                            & \multicolumn{2}{l|}{Class Description}                      &   Natural language functional description of the class \\ 
\multicolumn{1}{l|}{}                                                                         &                                                                            & \multicolumn{2}{l|}{Attributes}                             & Member variables with types, visibility, and constraints                   \\ 
\multicolumn{1}{l|}{}                                                                         &                                                                            & \multicolumn{2}{l|}{Relationships}                          & Inter-class connections                                                    \\ \cmidrule{3-5}
\multicolumn{1}{l|}{}                                                                         &                                                                            & \multicolumn{1}{l|}{\multirow{2}{*}{Methods}} & Signatures  & Method signatures with parameters, return types, and visibility       \\ 
\multicolumn{1}{l|}{}                                                                         &                                                                            & \multicolumn{1}{l|}{}                         & Description &  Natural language functional description of the function \\ \midrule
\multicolumn{2}{l|}{ \textit{Test Suite}}              & \multicolumn{2}{l|}{Test Cases}                    & Test cases to evaluate the correctness of the repository   \\ 
\bottomrule
\end{tabular}
\end{table}

\subsection{Benchmark Construction}
\begin{figure}[t] 
\centering
\includegraphics[width=\linewidth]{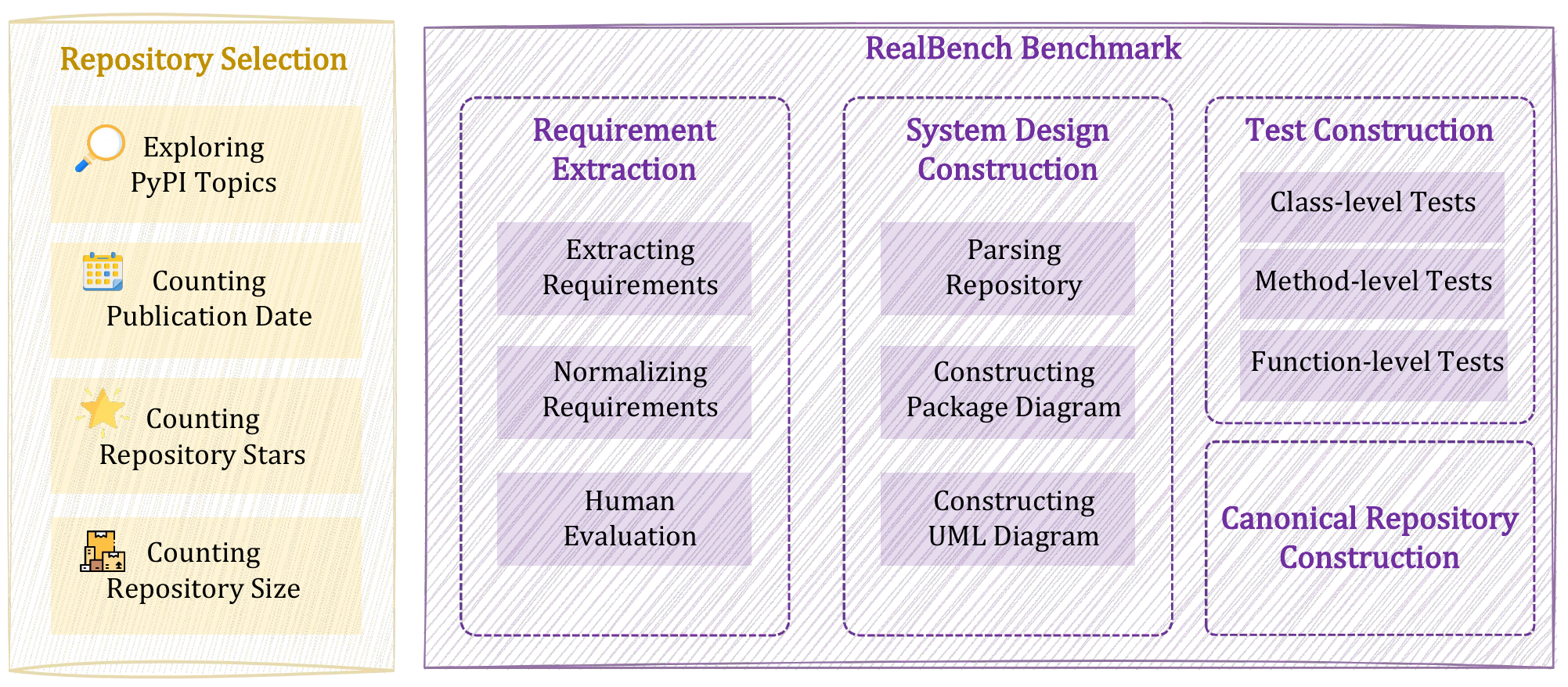}
\caption{The Construction Procedure of RealBench.}
\label{fig:workflow}
\end{figure}

Figure~\ref{fig:workflow} illustrates the construction procedure of RealBench. It mainly consists of four steps to create RealBench: (i) select suitable coding tasks (Section \ref{Task Selection}); (ii) construct requirements of repositories with natural language (Section \ref{Requirement Construction}); (iii) create system design based on the principles of a two-level standard UML diagram (Section \ref{System Design Construction}); and (iv) write the test suite for classes, methods, and functions within the repository (Section \ref{Test Creation}).

\subsubsection{Task Selection} \label{Task Selection}
Referring to the TIOBE index~\cite{tiobe-index} for programming language popularity, the most popular language is Python. Thus, we conduct experiments on Python and will expand to other languages in the future. We aim to ensure the representativeness of selected repositories in real-world programming scenarios. To this end, we focus on the following criterion:
(i) \textbf{\textit{Topic Selection}}: to enable RealBench covering more coding tasks in the real-world programming scenarios, we first identify the top-20 popular programming topics in PyPI\footnote{https://pypi.org/} since PyPI is the largest Python package registry; 
(ii) \textbf{\textit{Publication Date}}: to avoid as much the impact of data contamination, we select repositories that were published after 2024-12;
(iii) \textbf{\textit{Star Count}}: to ensure the quality of repositories, we require repositories to have more than 10 stars on GitHub;
(iv) \textbf{\textit{Project Sizes}}: to ensure RealBench covers a wide range of repository sizes, we divide repositories into four levels according to their lines of code (LOC), including 0$\sim$500 (level 1), 500$\sim$1000 (level 2), 1000$\sim$2000 (level 3), and $\geq$2000 (level 4).  
We crawl these repositories from GitHub and ensure they are non-forked and non-malicious. Finally, we obtain 61 real-world repositories.

\subsubsection{Requirement Construction} \label{Requirement Construction}

We manually construct the natural language requirements for each repository by inviting experienced Python programmers. Each participant has development experience at least 3 years. One participant is required to construct requirements, and the other is responsible for verifying and correcting the requirements. Specifically, if the README file contains requirements, one participant extracts requirement descriptions from the file. If a repository does not have requirements, the participant manually writes the requirements according to the code implementations. Then, the other participant manually verifies and corrects requirements, aiming to ensure the quality of our RealBench.

\subsubsection{System Design Construction} \label{System Design Construction}

We construct the system design for each repository with Understand\footnote{https://scitools.com/graphs}. Understand developed by SciTools is a software development tool designed to help programmers comprehend, maintain, and document their source code. 
Understand leverages its static code analysis engine to create and display UML diagrams such as class diagrams and package diagrams, particularly beneficial for understanding existing codebases and their architecture. These diagrams provide a visual overview of classes, their inheritance relationships, and the interactions and data flow between objects over time. 

During this step, we construct both the package diagram and class diagram for each repository. 
(i) \textit{\textbf{Package Diagram}} structures the high-level elements and dependencies in the system, which contains package names, public interfaces, and inter-package dependencies. We construct the package diagram following the UML standard~\cite{OMG2017UML}. Specifically, we first identify all the packages in the repository, along with their name, import statements, and public functions/classes. Then, based on the import statements, we extract the dependencies among the packages to form the final package diagram. 
(ii) \textit{\textbf{Class Diagram}} aims to reflect the detailed object-oriented design and static structure of the system, which contains complete class definitions with attributes, methods, and relationships. To construct it, we first identify all the classes as nodes in the diagram. Moreover, it is worth noting that Python does not strictly restrict all functions and variables to be contained in classes. Therefore, to allow the class diagram to reflect all the necessary details for the subsequent code generation task, we follow the practices to use service classes to encapsulate functions and variables that were not defined in a class. Next, we extracted all the function calls and data dependencies among the classes to identify the inter-class dependencies among the identified classes. 

To ensure the quality, we invite 3 participants to verify the correctness of the system design.
Participants have an average of 3 years of Python development experience. Each system design is initially assigned to two participants. They are required to check and rectify the system design, ensuring that it is consistent with the repository. In case of disagreements, a third participant facilitates discussions to reach a consensus on the system design.

\subsubsection{Test Creation} \label{Test Creation}

In this step, we manually create a test suite for each repository. Similar to system design construction, test suites are constructed by experienced programmers. One participant is responsible for creating the test cases, while the other verifies the correctness of the test cases. 
To conduct a comprehensive evaluation, we require participants to create test cases in three levels, including function-level, method-level, and class-level tests. 
\textbf{\textit{Function-level}} test evaluates a single function and only checks whether the return value is correct. 
\textbf{\textit{Method-level}} test focuses on independently verifying the correctness of each method within a class, where a method is invoked at a time without involving other methods. Different from the function-level test, it checks not only if the method's return value is correct but also inspects the method's impact on the class's fields. This test level can ensure that each method functions correctly on its own, without being impacted by other potentially incorrect methods.
\textbf{\textit{Class-level}} test aims to evaluate the overall functionality of the entire class, especially when multiple methods work together. It verifies if the interactions between methods are correct and whether the final state of the class and the outcome are as expected after a series of operations. In this step, we discarded projects that could not be executed to ensure the repositories in our dataset were testable.

To accelerate the test construction process, participants can use the existing test cases in repositories and also apply advanced LLMs (\ie GPT-4o) to help generate test cases. All tests are verified by the other participant.

\subsection{Benchmark Statistics}

In this way, we constructed our new benchmark RealBench. We summarize the main Statistics of RealBench in Table~\ref{tab: benchmark stastics}.

\textbf{Scale and Complexity.} Our benchmark contains a total of 61 repositories. The repositories are divided into four levels based on their size: Level-1 (0$\sim$500 LOC), Level-2 (500$\sim$1000 LOC), Level-3 (1000$\sim$2000 LOC), and Level-4 ($\geq$2000 LOC). As repository size increases, the complexity also grows significantly. Level-4 repositories contain the most complex structures with an average of 14.4 classes and 51.40 methods per repository.

\textbf{Dependency.} We analyze the percentage of standalone functions and methods (functions and methods with no dependencies on other code) at each level to measure task difficulty. On average, 44.73\% of methods and functions in RealBench are standalone. However, this proportion is much lower for large repositories: level-4 tasks contain only 26.23\% standalone methods and functions.

\textbf{Test Coverage and Quality.} Following the procedure described before, we manually create comprehensive test suites for each repository. On average, each repository contains 50.05 test cases. Our test coverage analysis shows that overall line coverage reaches 79.76\% across all repositories. Coverage varies by repository size, with smaller repositories achieving higher coverage rates. This pattern occurs because larger codebases are more complex, making complete coverage more challenging. Nevertheless, all coverage rates exceed 70\%, ensuring that our benchmark can effectively evaluate the correctness of generated code.

\begin{table}[t]
  \centering
  \caption{Statistics of RealBench (SA: Standalone, LC: Line Coverage).}
    \resizebox{\linewidth}{!}{
    \begin{tabular}{l|c|cccc|cc|ccc|c}
    \toprule
    \multirow{2}[2]{*}{} & \multicolumn{1}{c|}{\multirow{2}[2]{*}{LOC}} & \multicolumn{1}{c}{\multirow{2}[2]{*}{Class}} & \multicolumn{1}{c}{\multirow{2}[2]{*}{Method}}  & \multicolumn{1}{c}{\multirow{2}[2]{*}{Function}} &\multicolumn{1}{c|}{\multirow{2}[2]{*}{File}} & \multicolumn{2}{c|}{Dependency (\%)}  & \multicolumn{3}{c|}{Tests} & \multicolumn{1}{c}{\multirow{2}[2]{*}{Task}} \\
          &      &       &    &     &    &  SA   & Non-SA    & Num    & LOC  & LC (\%) \\
    \midrule
    Level-1 &  259.35 & 1.71  & 6.85  &  6.21  & 2.86   &  70.59  & 29.41  &   22.36    &  7.93   & 91.16 & 14 \\
    Level-2 & 845.88  & 8.29  & 23.71 &  11.71  &6.23  &  36.93   & 63.07  & 33.41  &  15.40  & 81.07  & 17 \\
    Level-3 & 1337.80  & 11.3  & 18.45    & 14.85  &  3.60  &  45.16   &  54.84 & 51.80  & 11.46     & 74.09 & 20  \\
    Level-4 &  2362.81  & 14.4  & 51.40    &  29.73   & 22.57   &  26.23  & 73.77  & 92.64  &  12.31 & 72.71 & 10 \\
    \midrule
    \#Avg   &  1201.46 &  8.92 &  25.10 &  15.63  &   8.82    &  44.73   &  55.27   &  50.05     &   11.78 & 79.76 & 15.25 \\
    \bottomrule
    \end{tabular}%
  \label{tab: benchmark stastics}%
  }
\end{table}%

%% file: sec3-Generation.tex
\section{Experiment Design}

In this section, we describe our experiments to evaluate LLMs' performance on our benchmark.
We first describe the code generation design using three generation strategies (Section \ref{Studies Generation Strategies}) and two system design formats (Section \ref{UML Diagrams Ablation}). Then, we present a systematic evaluation design for assessing the generated repositories, which contains two evaluation granularities (Section \ref{Evaluation Granularity}) with five metrics (Section \ref{Evaluation Metrics}). Next, we provide the prompt design used in the generation process (Section \ref{Prompt Design}), studied LLMs (Section \ref{Studied Large Language Models}), and implementation details (Section \ref{Implementation Details}).

\subsection{Code Generation Design} \label{Code Generation Design}

Given natural language requirements and a UML diagram, we generate an entire repository class-by-class, composing file-by-file. In a repository, code components usually have complex dependencies, such as import relations from one file to another. Thus, we study three generation strategies (Section \ref{Studies Generation Strategies}). Since the system design formats also matter, we also investigate two types of system designs (Section \ref{UML Diagrams Ablation}). 

\begin{figure}[t] 
\centering
\includegraphics[width=\linewidth]{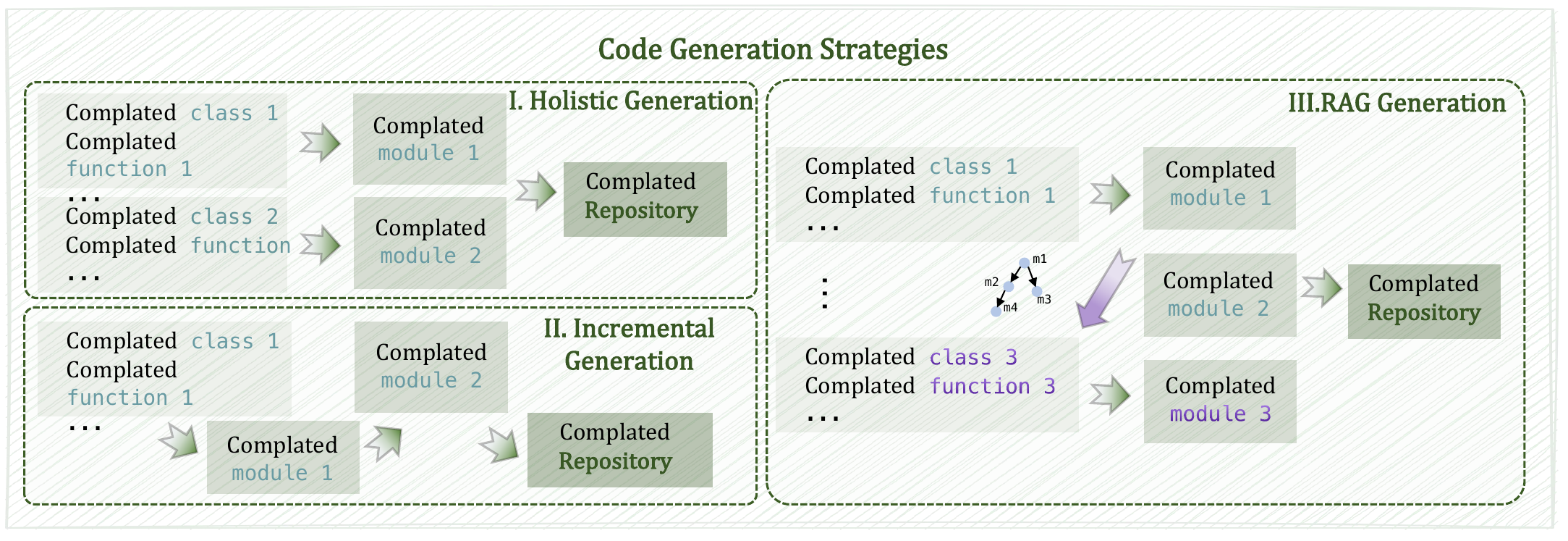}
\caption{Illustration of Three Code Generation Strategies for Each Repository.}
\label{fig: generation workflow}
\end{figure}

\subsubsection{Generation Strategies}  \label{Studies Generation Strategies}

We design three repo-level code generation strategies as shown in Figure~\ref{fig: generation workflow}. A straightforward setting asks LLMs to generate the entire repository all at once with the natural language requirements and the system design, named \textbf{\textit{Holistic Generation}}. Considering that LLMs struggle to maintain long-range coherence in long-form generations \cite{ye2025longproc}, we investigate \textbf{\textit{Incremental Generation}}. In this setting, LLMs generate code in a file-by-file manner to reduce the generated context length. Each iteration is based on requirements and the system design. The iterative process repeats until all classes are generated. Code components in a repository usually have dependencies, \eg one file imports another. Inspired by this, we design the \textbf{\textit{Retrieval-Augmented Generation}} setting. 
It requires LLMs to generate one file each time based on the retrieved relevant files generated in previous iterations, along with the requirements and the system design. To retrieve relevant files, we construct a directed acyclic graph to describe the relationships between files according to the system design. Then, we select files that have the ``import'' relation with the target file. This strategy keeps a balance between rich domain knowledge and limited context windows of LLMs.

\begin{figure}[t] 
\centering
\includegraphics[width=\linewidth]{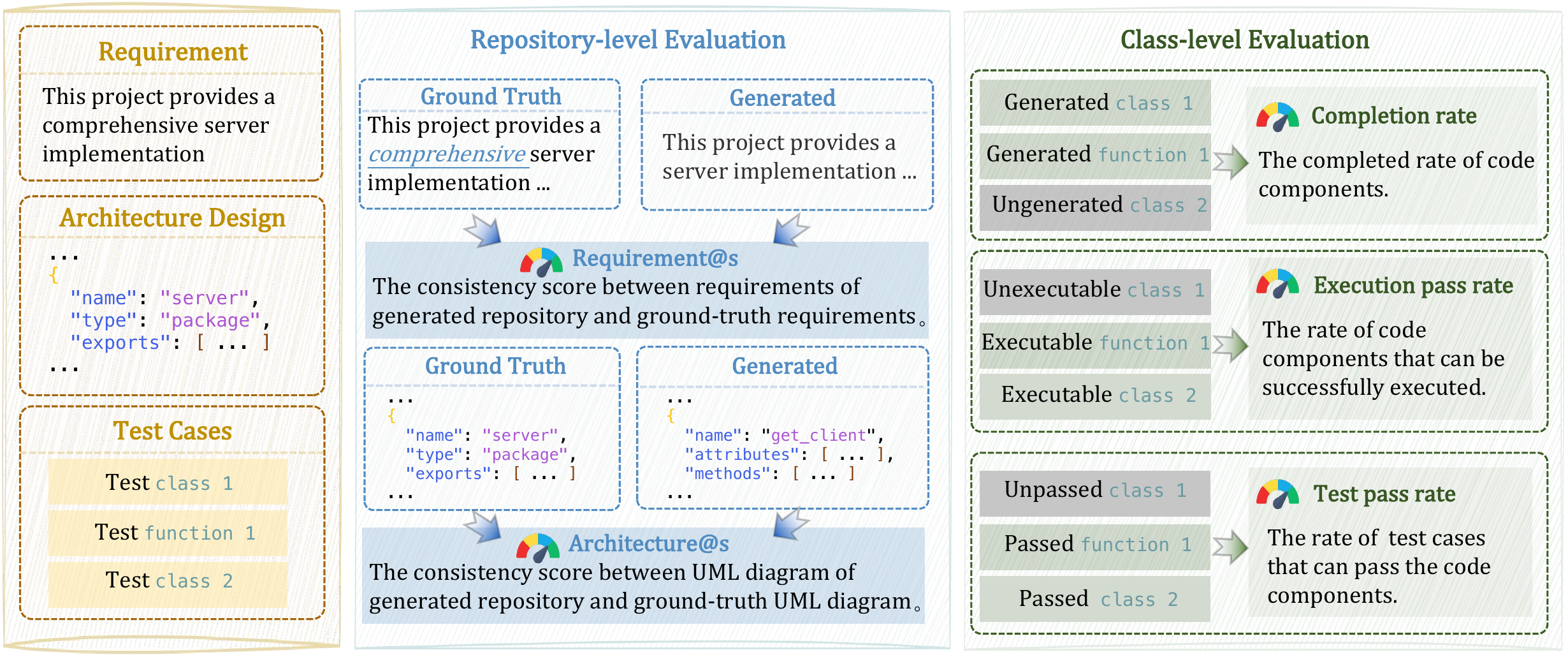}
\caption{Evaluation design for assessing the generated repository. It contains repository-level and class-level granularities with five metrics. Repository-level metrics evaluate whether the generated repository satisfies the ground-truth requirements and the system design. Class-level metrics assess whether the code components of generated repositories are completed, executed, and passed test cases.}
\label{fig: evaluation-workflow}
\end{figure}

\subsubsection{System Design Types} \label{UML Diagrams Ablation}
We investigate two system design methods for code generation. The first method uses only UML package diagrams, which show packages and their dependencies. This provides a high-level view of the system architecture.
The second method combines UML package diagrams with class diagrams, where class diagrams show class names, attributes, methods, and relationships. These relationships include inheritance, associations, and dependencies between classes. This method provides both high-level and low-level architectural designs.

We chose these two methods based on different industry software development methodologies. For agile development, teams prefer minimal documentation and design artifacts. Package diagrams provide a good balance. They show the high-level system structure without too many implementation details. For waterfall development, teams need complete upfront design and documentation. Therefore, we use both package diagrams and detailed class diagrams to simulate this scenario: Package diagrams show the overall structure, while class diagrams add specific implementation details.

\subsection{Evaluation Design} 

To evaluate the quality of generated repositories, we use two sets of metrics. The first set includes test-based metrics (\ie Completion rate, Execution pass rate, and Test pass rate) that evaluate behavioral correctness \cite{cao2024javabench}. However, these metrics only assess functional correctness while ignoring non-functional aspects like performance or security design. To address this limitation, we introduce two repository-level metrics (\ie Requirement@$k$ and Architecture@$k$) that evaluate whether the generated code matches the original requirements and design.

\subsubsection{Evaluation Granularity} \label{Evaluation Granularity}

We design two evaluation granularities (\ie repository-level and class-level) to comprehensively assess a repository. Repository-level granularity evaluates whether the generated repository meets the ground-truth requirements and the system design. Class-level granularity evaluates whether the code components (\ie functions, classes, and methods) within generated repositories are completed, executed, and passed test cases. 

\textbf{\textit{Repository-Level}} granularity evaluates whether the generated repository meets the users' requirements and the system design. 
We employ DeepWiki\footnote{https://deepwiki.com/}, an advanced repository-understanding code agent, to automatically analyze generated repositories. Given a repository, DeepWiki can analyze it from diverse aspects such as generating its requirements, system design, and environment setup, while also supporting multi-turn conversational interactions. 
To evaluate a generated repository, we first use DeepWiki to generate its requirement descriptions and a class diagram. Subsequently, one participant with 3 years of programming experience verifies and corrects the output, ensuring requirements and system design are consistent with the generated repository. Another participant then compares the validated descriptions and class diagram against the ground-truth labels provided by our benchmark. Finally, they assign a score (\ie [0, 4]) based on our designed comparison criteria, where the detailed instructions are available on our website.
The higher the score is, the better the quality of the generated repository will be.

To conduct a fine-grained evaluation, we design \textbf{\textit{Class-Level}} to assess the code components (\ie functions, classes, and methods) within repositories, thereby facilitating a deeper analysis of LLMs' abilities in repo-level code generation. We evaluate code components at three progressive levels, including completeness, executability, and correctness. Completeness measures the proportion of code components that are generated, which reflects whether LLMs can identify code elements in the system design. Executability assesses the percentage of code elements that can be successfully run. It focuses on evaluating the syntax correctness of generated repositories. 
Correctness, the most stringent evaluation design, measures the functional accuracy of code elements within repositories. This criterion not only requires the code to be runnable and syntactically correct, but also demands that its functionality is accurate.

\subsubsection{Evaluation Metrics} \label{Evaluation Metrics}

We design five metrics for repository-level and class-level evaluation.
For repository-level evaluation, we design \textbf{\textit{Requirement@k}} and \textbf{\textit{Architecture@k}} metrics, which are human evaluation scores by comparing requirement descriptions and system designs of generated repositories against the ground-truth labels, ranging from 0 to 4.  

For class-level evaluation, we adapt \textbf{\textit{Test pass rate}} \cite{chen2021evaluating} computed as the number of passed tests divided by the total number of tests for each task.
Similar to \textbf{\textit{Test pass rate}}, we design Completion rate and Execution pass rate. \textbf{\textit{Completion rate}} measures the expectation of completed code components (\ie class, function, and method) in a generated repository. \textbf{\textit{Execution pass rate}} is the expectation of code components that can be successfully executed.

\subsection{Prompt Design} \label{Prompt Design}

Following the common practice of prompting LLMs \cite{cao2024javabench, du2023classeval}, we design the prompt used for the repo-level code generation aligned with real-world software development practice. The prompt contains two parts: a system prompt as the beginning instruction to initialize LLMs, and a task instruction to describe the purpose of the task. Each generation strategy is defined in the task instruction. The designed prompt is shown as follows:

\noindent 
\begin{tcolorbox}
\textbf{System Prompt:} You are a senior Python programmer familiar with Python programming and UML diagrams.
\end{tcolorbox}

\vspace{1em} 

\noindent
\begin{tcolorbox}
\textbf{Instruction:}  Generate an entire Python project based on the given natural language requirements and UML diagram of the project.
The generated output format meeting the following demands:
(i) Each module starts with <begin$\_$module> and ends with <end$\_$module>; 
(ii) Adds necessary imports between modules.

\vspace{1em}

The given natural language requirements:

\texttt{\$\{Project$\_$Requirements\}}

\vspace{1em}

The given UML diagram with JSON format:

\texttt{\$\{UML$\_$JSON$\_$String\}}

\vspace{1em}
Do not include any explanations or additional text.
\end{tcolorbox}

\subsection{Studied Large Language Models} \label{Studied Large Language Models}

We evaluate 6 advanced LLMs on RealBench, which contain 5 general models (\ie GPT-4o \cite{GPT-4o}, Gemini 2.5 Flash \cite{Gemini-2.5-flash}, Claude Sonnet 4 \cite{Claude-Sonnet-4}, Deepseek-V3 \cite{guo2024deepseek} and Qwen3-235B-A22B \cite{Qwen3-235B-A22B}) and 1 code model (Qwen2.5-Coder-7B-Instruct~\cite{Qwen2.5-Coder-7B}). We considered the instruction version of LLMs because we need to use the instruction-following ability. Table \ref{tab: llms} shows the 6 LLMs studied in this paper with their releasing time (Column ``Time''), model sizes (Column ``Size''), and context window lengths (Column ``Window Length'').

\begin{table}[htbp]
  \centering
  \caption{ Studied Large Language Models}
      \resizebox{0.7\linewidth}{!}{
    \begin{tabular}{l|l|ccc}
    \toprule
    \multicolumn{1}{l}{ Model} & \multicolumn{1}{l}{Type} & \multicolumn{1}{c}{Size} & \multicolumn{1}{c}{Time} & \multicolumn{1}{c}{Window Length} \\
    \midrule
    \multirow{3}[2]{*}{General Model} &  GPT-4o \cite{GPT-4o}      &  --  &   May, 24    & 128K \\
      & Claude-Sonnet-4 \cite{Claude-Sonnet-4}   &   --    &     2025-05      & 200K \\
    &    Gemini-2.5-Flash \cite{Gemini-2.5-flash}         & -- &    2025-04   & 1,048K \\
    \midrule
    \multirow{3}[2]{*}{Open Source Model}
    &  Deepseek-V3 \cite{liu2024deepseek}       &   671B    &    2024-12   & 64K \\
     &  Qwen3-235B-A22B \cite{Qwen3-235B-A22B}       &  235B &   2025-04    & 128K \\
     &  Qwen2.5-Coder-7B-Instruct \cite{Qwen2.5-Coder-7B}       & 7B &    2024-11   & 128K \\
    \bottomrule
    \end{tabular}%
  \label{tab: llms}%
  }
\end{table}%

\subsection{Implementation Details} \label{Implementation Details}

We use greedy search for all experiments. For each programming task, LLMs generate one solution. Constrained by model availability and computing resources, we evaluate all studied LLMs by invoking APIs provided by their official interfaces. For instance, we use the OpenAI API interface for using GPT-4o (\ie gpt-4o-2024-05-13) and interfaces provided by Anthropic for invoking claude-sonnet-4-20250514.
Although LLMs support context windows exceeding hundreds of thousands of tokens, the input length of retrieval-augmented generation strategies sometimes exceeds LLMs' context windows. We truncate the retrieved Python files from the front of the code sequence. The invited developers in evaluation are PhD candidates majoring in computer science with programming experience for at least three years.

%% file: sec4-experiments.tex
\section{Evaluation}

We conduct extensive studies to evaluate LLMs' capabilities on repo-level code generation aligned with real-world software development practice.  
The research questions (RQs) are designed as follows:

\begin{itemize}[leftmargin=*]
    \item \textbf{RQ1: Overall Performance}. We evaluate the performance of the studied LLMs on RealBench. We use our three code generation strategies with UML diagram and its ablation version (Section \ref{UML Diagrams Ablation}) to generate repositories.
    \item \textbf{RQ2: Generation Strategies}. We investigate different generation strategies (Section \ref{Studies Generation Strategies})  and analyze their influences on repo-level code generation. 
    \item \textbf{RQ3: UML Diagram Ablation}. The UML diagram contains complicated components. We explore the impacts of different components (Section \ref{UML Diagrams Ablation}).
    \item \textbf{RQ4: Bad Case Analysis}. We analyze the common errors and provide insights for LLMs in repo-level code generation aligned with software development practice. 
\end{itemize}

\begin{table}[tbp]
  \centering
  \caption{Overall Performance of LLMs on RealBench.}
  \resizebox{\linewidth}{!}{
    \begin{tabular}{l|l|cccc|cccc|cccc|cccc|cccc}
    \toprule
    \multirow{3}[4]{*}{Strategies} & \multirow{3}[4]{*}{Models} & \multicolumn{12}{c|}{Class-level} & \multicolumn{8}{c}{Repository-level} \\
\cmidrule{3-22}          &       & \multicolumn{4}{c|}{\textit{Completion rate}}        & \multicolumn{4}{c|}{\textit{Execution pass rate}}        & \multicolumn{4}{c|}{\textit{Test pass rate}} & \multicolumn{4}{c|}{\textit{Requirement@1}}       & \multicolumn{4}{c}{\textit{Architecture@1}}       \\
          &       & \multicolumn{1}{c}{L1} & \multicolumn{1}{c}{L2} & \multicolumn{1}{c}{L3} & \multicolumn{1}{c|}{L4} & \multicolumn{1}{c}{L1} & \multicolumn{1}{c}{L2} & \multicolumn{1}{c}{L3} & \multicolumn{1}{c|}{L4} & \multicolumn{1}{c}{L1} & \multicolumn{1}{c}{L2} & \multicolumn{1}{c}{L3} & \multicolumn{1}{c|}{L4} & \multicolumn{1}{c}{L1} & \multicolumn{1}{c}{L2} & \multicolumn{1}{c}{L3} & \multicolumn{1}{c|}{L4} & \multicolumn{1}{c}{L1} & \multicolumn{1}{c}{L2} & \multicolumn{1}{c}{L3} & \multicolumn{1}{c}{L4} \\
    \midrule
    \multicolumn{1}{l|}{\multirow{7}[4]{*}{Holistic}} 
    & GPT-4o    & \colorcell{96.97}    & \colorcell{87.60} & \colorcell{52.56}   & \colorcell{92.15}  &\colorcell{90.73} & \colorcell{38.26}  & \colorcell{40.80} & \colorcell{81.45} &\colorcell{24.28}  &\colorcell{5.39} &\colorcell{19.48} &\colorcell{13.51} & \colorcell{2.63}   & \colorcell{1.42}  & \colorcell{2.32}  &\colorcell{1.80}         & \colorcell{3.14} & \colorcell{3.48} & \colorcell{2.47}  & \colorcell{1.60}  \\
    & Claude-Sonnet-4 & \colorcell{92.12}  & \colorcell{85.17} & \colorcell{53.24}   & \colorcell{90.21}  &\colorcell{83.29} & \colorcell{34.31}  & \colorcell{35.09} & \colorcell{72.15} &\colorcell{28.65}  &\colorcell{7.34} &\colorcell{16.18} &\colorcell{14.30} & \colorcell{2.72}   & \colorcell{0.89}  & \colorcell{1.54}  &\colorcell{1.70}         & \colorcell{2.87} & \colorcell{1.75} & \colorcell{1.87}  & \colorcell{0.60}  \\
     & Gemini-2.5-Flash  & \colorcell{49.70}  & \colorcell{60.11}  &\colorcell{60.03} &\colorcell{78.95}  &\colorcell{23.52}  & \colorcell{16.51} & \colorcell{16.42} & \colorcell{24.55} & \colorcell{15.65} &\colorcell{5.95} & \colorcell{6.70} & \colorcell{3.07} & 2.04  &  0.72   & 0.74 & 1.10 & 1.85 & 1.63 & 1.02 & 1.70 \\
     & Deepseek-V3    & \colorcell{89.70}  &\colorcell{74.12} &\colorcell{32.63} &\colorcell{44.82}  &\colorcell{77.64}  & \colorcell{21.95} & \colorcell{27.60} & \colorcell{4.46} & \colorcell{43.13}  & \colorcell{11.23}  & \colorcell{21.00}  & \colorcell{2.21} & 3.17  & 1.75  &2.26  & 0.60  & 2.61  & 1.71  & 2.69  & 1.70  \\
     & Qwen3-235B-A22B   & \colorcell{31.52} & \colorcell{52.90}  & \colorcell{69.28} &\colorcell{38.45}  & \colorcell{16.42}  &\colorcell{10.84}  & \colorcell{37.21}  & \colorcell{5.48} & \colorcell{19.81} & \colorcell{6.74}  & \colorcell{18.34}  & \colorcell{0.91}  & 2.45 & 1.38  &  2.81  & 0.80 & 1.82 & 0.74  &2.80   & 0.60   \\
    & Qwen2.5-Coder-7B     & \colorcell{34.60}  & \colorcell{54.38}  & \colorcell{63.41} & \colorcell{41.37}  & \colorcell{57.31} &\colorcell{19.43} & \colorcell{49.42} & \colorcell{12.06}  &  \colorcell{21.36}   & \colorcell{7.13}   & \colorcell{16.57}  &\colorcell{0.85} &  2.61  &1.53  & 2.85  & 0.7   & 2.07  &  0.92  & 2.67   & 0.80  \\
\cmidrule{2-22}          & Average   & \colorcell{65.77} & \colorcell{69.05} & \colorcell{55.19} & \colorcell{64.33} & \colorcell{58.15} & \colorcell{23.55} & \colorcell{34.42} & \colorcell{33.36} & \colorcell{25.48} & \colorcell{7.30} & \colorcell{16.38} & \colorcell{5.81} & \colorcell{2.60} & \colorcell{1.28} & \colorcell{2.09} & \colorcell{1.12} & \colorcell{2.39} & \colorcell{1.71} & \colorcell{2.25} & \colorcell{1.17} \\
    \midrule
    \multicolumn{1}{l|}{\multirow{7}[4]{*}{Incremental}}
    & GPT-4o    & \colorcell{100.00}  &  \colorcell{87.47}  &\colorcell{87.02}  &\colorcell{90.26}  & \colorcell{84.66} &\colorcell{44.81} &\colorcell{58.32} &\colorcell{48.69}  & \colorcell{20.45} &\colorcell{6.21} & \colorcell{23.44} & \colorcell{15.48} & 0.92  &  1.20  & 2.21  &  1.70  & 2.39 & 2.12  & 2.37  & 2.10   \\
     & Claude-Sonnet-4    & \colorcell{92.73}  &  \colorcell{78.98}  &\colorcell{82.90}  &\colorcell{84.61}  & \colorcell{92.73} &\colorcell{46.24} &\colorcell{56.15} &\colorcell{57.49}  & \colorcell{26.80} &\colorcell{10.75} & \colorcell{17.37} & \colorcell{15.11} & 0.87  &  1.20  & 2.21  &2.00 &  2.94  & 1.45 & 1.82  &1.70    \\
          & Gemini-2.5-Flash     & \colorcell{100.00} & \colorcell{88.81} &\colorcell{86.76}  &\colorcell{86.39} &\colorcell{69.97}  &\colorcell{63.55} &\colorcell{48.74}  & \colorcell{42.08}  & \colorcell{5.64} & \colorcell{3.68} &\colorcell{19.03} & \colorcell{12.04} & 0.25  & 0.37   & 2.10  & 1.40  &  1.03  & 1.14   &2.03  & 1.40   \\
          & Deepseek-V3  & \colorcell{91.31} & \colorcell{77.35}  &\colorcell{34.19} &\colorcell{49.35}  & \colorcell{79.08}  & \colorcell{39.66} &\colorcell{20.65} &\colorcell{21.20}  & \colorcell{17.18}  & \colorcell{8.09}   &\colorcell{23.44}   & \colorcell{14.62} & 0.97  & 1.29  & 2.23   & 1.60  &  1.41     &  1.78   & 2.58   & 1.30   \\
     & Qwen3-235B-A22B   & \colorcell{96.97}  &\colorcell{85.98} &\colorcell{77.59}  & \colorcell{78.01}  & \colorcell{83.21}  &\colorcell{27.86} &\colorcell{24.60}  &\colorcell{24.74} &\colorcell{31.95}  &\colorcell{4.09}  &\colorcell{31.51} &\colorcell{10.32} & 2.74  & 0.43  & 2.49  & 1.10  & 2.82  &1.60 &  2.81  & 1.50  \\
     & Qwen2.5-Coder-7B  & \colorcell{93.05} & \colorcell{84.72} &\colorcell{80.35} &\colorcell{81.43}  & \colorcell{84.10}  &  \colorcell{36.47}  & \colorcell{47.65}  & \colorcell{35.07}  & \colorcell{28.71}  &\colorcell{5.16}  &\colorcell{28.47}  & \colorcell{12.56}  & 2.83  & 0.94  & 2.75  &  1.20  & 2.97  &1.33 &  2.76 & 1.30   \\
\cmidrule{2-22}          & Average   & \colorcell{95.68} & \colorcell{83.89} & \colorcell{74.80} & \colorcell{78.43} & \colorcell{82.29} & \colorcell{43.10} & \colorcell{42.69} & \colorcell{38.21} & \colorcell{21.79} & \colorcell{6.33} & \colorcell{23.88} & \colorcell{13.36} & \colorcell{1.43} & \colorcell{0.91} & \colorcell{2.33} & \colorcell{1.50} & \colorcell{2.26} & \colorcell{1.57} & \colorcell{2.40} & \colorcell{1.55} \\
    \midrule
    \multicolumn{1}{l|}{\multirow{7}[4]{*}{RAG}} 
    & GPT-4o    &  \colorcell{89.09}  &  \colorcell{80.73}  &   \colorcell{82.83}  &  \colorcell{91.83} & \colorcell{82.54} & \colorcell{40.07}   & \colorcell{45.67} &  \colorcell{46.37}  & \colorcell{24.60} & \colorcell{5.73}  &\colorcell{19.10}  & \colorcell{9.95} & 2.71  & 1.28 & 1.41   & 1.40   & 2.27   &    1.43  &2.56 &  1.10  \\
     & Claude-Sonnet-4   &  \colorcell{86.97}  &  \colorcell{75.56}  &   \colorcell{78.39}  &  \colorcell{86.34} & \colorcell{79.55} & \colorcell{40.33}   & \colorcell{55.89} &  \colorcell{52.55}  & \colorcell{26.52} & \colorcell{5.07}  &\colorcell{19.49}  & \colorcell{21.33} & 2.43  & 1.05 & 2.37   & 1.80   & 2.37  &  0.71  &2.50  &  2.80  \\
          & Gemini-2.5-Flash    & \colorcell{72.04} & \colorcell{78.25}  &\colorcell{72.05} & \colorcell{65.30} &   \colorcell{56.94}   & \colorcell{30.47}  & \colorcell{27.16}   & \colorcell{15.92}  & \colorcell{19.40}  & \colorcell{4.54}   &  \colorcell{14.91}   & \colorcell{8.01}  &  \colorcell{1.62} & \colorcell{1.47}   & \colorcell{1.96}  &  \colorcell{1.30}   &  \colorcell{3.26}  & \colorcell{1.10}   & \colorcell{2.31}  & \colorcell{1.30}   \\
          & Deepseek-V3    &   \colorcell{88.48}  & \colorcell{74.66}  & \colorcell{50.72}   &\colorcell{74.66}   & \colorcell{64.68}  & \colorcell{24.17}   & \colorcell{28.48} & \colorcell{37.14} & \colorcell{41.85}  &\colorcell{5.03}  &\colorcell{17.37}  &  \colorcell{13.27} & 3.35  &  0.97  &1.67  & 1.40 &  3.03 &1.21  & 1.24  & 1.70  \\
          & Qwen3-235B-A22B   & \colorcell{80.61}  & \colorcell{74.35}  & \colorcell{60.03} & \colorcell{68.06} & \colorcell{64.68}  &   \colorcell{19.72}    &   \colorcell{20.93}    &   \colorcell{15.17}    &  \colorcell{9.23} & \colorcell{4.12} & \colorcell{7.16} & \colorcell{0.73}      & 1.72  & 0.54   & 1.20 & 0.5  &1.82  &  0.93   & 1.06  &   0.80    \\
          & Qwen2.5-Coder-7B     &  \colorcell{31.52}   & \colorcell{46.77}  &\colorcell{30.28}  &\colorcell{20.63}   & \colorcell{22.26}  & \colorcell{11.95} &\colorcell{15.91} &   \colorcell{11.30}   & \colorcell{8.95}    & \colorcell{9.49}  & \colorcell{17.08} & \colorcell{15.11} & 1.29 & 0.72  & 0.93  & 1.30  &0.74  & 0.56  & 0.79   & 1.80   \\
\cmidrule{2-22}          & Average   & \colorcell{74.79} & \colorcell{71.72} & \colorcell{62.38} & \colorcell{67.80} & \colorcell{61.78} & \colorcell{27.79} & \colorcell{32.34} & \colorcell{29.74} & \colorcell{21.76} & \colorcell{5.66} & \colorcell{15.85} & \colorcell{11.40} & \colorcell{2.19} & \colorcell{1.01} & \colorcell{1.59} & \colorcell{1.28} & \colorcell{2.25} & \colorcell{0.99} & \colorcell{1.74} & \colorcell{1.58} \\
    \bottomrule
    \end{tabular}%
  \label{tab:main results}%
  }
\end{table}%

\subsection{RQ1: Overall Performance}

Table \ref{tab:main results} presents the performance of LLMs on RealBench with greedy sampling. We have the following observations.

\textbf{Comparison among Models.}
We observe significant performance gaps among LLMs. Among open source models, DeepSeek-V3 (671B) performs the best, which is related to its large number of parameters. For Qwen series models, Qwen3-235B-A22B achieves similar performance to Qwen2.5-Code-7B in most scenarios, despite the significant differences in parameter sizes. For example,  Qwen3-235B-A22B achieves 19.81\% test pass rate and Qwen2.5-Code-7B performs 21.36\% test pass rate on Level-1 in holistic strategy. The reason might be that Qwen2.5-Code-7B is a code model specifically designed for solving programming tasks. 
In proprietary source models, GPT-4o and Claude-Sonnet-4 significantly outperform Gemini-2.5-Flash, where Gemini-2.5-Flash achieves unsatisfactory performance. 
Generally, the performances of studied LLMs are no more than 20\% test pass rate in most experiments, verifying the limited abilities of LLMs on real-world repo-level code generation.

\vspace{1mm}
\begin{custommdframed}
\textbf{Finding 1:} The best average completion rate, Execution pass rate, Test pass rate, Requirement@1, and Architecture@1 scores are 91.18\%, 45.00\%, 19.39\%, 2.29, and 2.67 on RealBench among the studied LLMs. This demonstrates that LLMs have limited capabilities on real-world repo-level code generation. Additionally, there are significant performance gaps among LLMs. 
\end{custommdframed}
\vspace{0mm}

\textbf{Performance on Different Metrics.}
In class-level metrics, completion rate scores are always higher than execution pass rate, and test pass rate scores are the lowest among the three metrics. This is in line with the fact that test pass rate is the strictest metric, which not only requires generated code to be runnable but also has the correct functionality. For completion assessment, we can find that most LLMs achieve high completion rates even on difficulty repositories (\ie Level-4). For instance, completion rate scores of GPT-4o are more than 80\% over all levels. This reflects that LLMs have satisfactory abilities to understand system design and can effectively identify modules within diagrams. 
For execution evaluation, LLMs can achieve decent performance on simple repositories (\ie Level-1). With the complexity of repositories increasing, execution pass rate drops dramatically (\eg from 82.29\% (Level-1) to 38.21\% (Level-4) on the incremental strategy). Meanwhile, compared to completion rate, execution scores are significantly lower. The above phenomena show that even if LLMs can generate code for modules in system designs, the generated code has non-negligible execution errors. 
We observe a significant decrease in test pass rate compared to the other two metrics in fine-grained class-level. In particular, LLMs achieve around 0\%$\sim$20\% test pass rate scores in most scenarios, which shows substantially low performance of   LLMs on repo-level code generation.
From the results on repository-level metrics, we can find that the generated repositories can not completely implement users' requirements and fail to maintain consistency with ground-truth system designs in most scenarios.

\vspace{1mm}
\begin{custommdframed}
\textbf{Finding 2:} Our designed metrics can effectively reflect qualities of generated repositories in finer granularities. LLMs can achieve satisfactory average completion rates (> 50\%). However, LLMs on execution pass rate and test pass rate achieve lower performances. 
This shows that while LLMs are good at finding and creating modules defined in UML diagrams, the quality of generated modules is often poor due to grammar and logic errors.
\end{custommdframed}
\vspace{0mm}

\textbf{Empirical Lessons.} If only using open source models locally and having limited computing resources, we suggest users to select Qwen2.5-Coder-7B, which can achieve similar performances with the larger open source model Qwen3-235B-A22B. Among proprietary source models, we recommend users to apply GPT-4o or Claude-Sonnet-4, instead of Gemini-2.5-Flash.

\subsection{RQ2: Generation Strategies}

We compare the performance of three generation strategies (\ie holistic, incremental, and RAG) as shown in Figure \ref{tab:main results}. We find that the best generation strategy varies with the difficulty levels of repositories.   

\textbf{Comparison across Different Levels.} Holistic generation strategy achieves the best performance on easy repositories (\ie level-1 and level-2 repositories) with 25.48\% and 7.30\% in test pass rate. Incremental strategy is the best generation strategy on level-3 and level-4 repositories, which performs much higher on test pass rate than holistic and retrieval-augmented strategies (23.88\% and 13.36\% average improvements in test pass rate). 
For example, on level-1 and level-2 repositories, DeepSeek-V3 achieves 43.13\% and 11.23\% test pass rate in holistic strategy, while performing 11.18\% and 8.09\% test pass rate in incremental generation. On level-3 and level-4 examples, DeepSeek-V3 performs 21.00\% and 2.21\% test pass rate in holistic strategy, and achieves 23.44\% and 14.62\% test pass rate in incremental generation. This reflects that generating a complex repository at once is more challenging for LLMs compared to generating each module in separate iterations, which might be explained by the limited long code generation abilities of LLMs as observed in the study \cite{ye2025longproc}.

\vspace{1mm}
\begin{custommdframed}
\textbf{Finding 3:} Generating the entire repository at once (\ie holistic generation strategy) is the best generation strategy on smaller repositories (\ie level-1 and level-2 repositories). For generating complex repositories (level-3 and level-4), the incremental strategy works better. This might be attributed to the limited ability of LLMs' long code generation abilities.
\end{custommdframed}
\vspace{0mm}

\textbf{Comparison between Holistic and Incremental Strategies.} 
The performance of holistic generation is better than RAG strategy in generating repositories. RAG achieves 2.67\% average decline compared to incremental strategy in test pass rate over level-1 to level-4. The performance trend in execution pass rate is the same as the counterpart in test pass rate. Execution pass rate descends from 51.57\% on incremental generation to 37.91\% on RAG strategy. We can find that providing the retrieved code actually hinders generating the correct code. As described in Section \ref{Studies Generation Strategies}, RAG strategy requires LLMs to generate each module at once based on the retrieved relevant modules generated in previous iterations. One potential reason for the performance above might be that retrieved code contributes to the long input contexts. It is more challenging for LLMs to fully understand and utilize the long input since LLMs have limited capabilities of understanding and utilizing long input contexts, as observed in the study \cite{li2025longcodeu}. In addition to the limited capability of handling long inputs mentioned above, we analyze the generated repositories in RAG strategy and find that LLMs prefer to combine generated code and retrieved code elements into a single module, failing to follow module relationships in diagrams and leading to errors.

\vspace{1mm}
\begin{custommdframed}
\textbf{Finding 4:} Incremental generation performs better than RAG strategy. 
The poor performance of RAG strategy may come from LLMs' limited ability to understand long code. Also, LLMs often combine generated code and retrieved code into a single module, which changes the module relationships shown in the original diagrams.
\end{custommdframed}
\vspace{0mm}

\textbf{Empirical Lessons.}
Based on the above experiments, we summarize the empirical lessons we learned as: Choosing holistic strategy when users' requirements are easy (\ie the line of target repositories is less than 1000). Otherwise, incremental strategy is recommended to generate complex repositories.

\subsection{RQ3: System Design Ablation}

In RQ1 and RQ2, the system design is set to the combination of UML class diagrams with package diagrams, where package diagrams show the overall structure while class diagrams add specific implementation details. In this RQ, we investigate the impact of diagrams by removing class diagrams.  Table \ref{tab:ablation} presents the performance of LLMs on RealBench when only accessing to high-level package diagrams. 

By comparing Table \ref{tab:main results} and \ref{tab:ablation}, we can make three observations. First, only providing high-level architectural designs is insufficient without implementation details. The test pass rate scores drop dramatically, even reaching zero in some scenarios, meaning that it is almost impossible to generate repo-level code that can pass test cases. For instance, a 12.74\% drop (16.34\% - 3.60\%) of test pass rate in the incremental strategy can be observed. 
Second, removing class diagrams significantly impacts the execution pass rate of generated repositories. For example, execution pass rate drops from 37.37\% to 23.01\% in holistic strategy. We argue that the execution failure might be caused by the absence of implementation details such as class attributes and methods. 
Third, it is clear that introducing system designs to LLMs is non-negligible in repo-level code generation tasks. Since removing the architecture information leads to a performance drop. The reason is that unlike generating standalone statement-level or function-level code, repo-level code generation is a complex programming task, which needs system designs to model modules, interfaces, and their relationships.

\vspace{1mm}
\begin{custommdframed}
\textbf{Finding 5:} Generating repositories with both UML package diagrams (high-level) and class (low-level) diagrams can yield better performance compared to providing only package diagrams. High-level and low-level system structures both benefit LLMs in generating correct code.
\end{custommdframed}
\vspace{0mm}

\begin{table}[tbp]
  \centering
  \caption{Performance of LLMs without Class Diagram.}
  \resizebox{\linewidth}{!}{
    \begin{tabular}{l|l|cccc|cccc|cccc|cccc|cccc}
    \toprule
    \multirow{3}[4]{*}{Strategies} & \multirow{3}[4]{*}{Models} & \multicolumn{12}{c|}{Class-level } & \multicolumn{8}{c}{Repository-level } \\
\cmidrule{3-22}          &       & \multicolumn{4}{c|}{\textit{Completion rate}}        & \multicolumn{4}{c|}{\textit{Execution pass rate}}        & \multicolumn{4}{c|}{\textit{Test pass rate}} & \multicolumn{4}{c|}{\textit{Requirement@1}}       & \multicolumn{4}{c}{\textit{Architecture@1}}       \\
          &       & \multicolumn{1}{c}{L1} & \multicolumn{1}{c}{L2} & \multicolumn{1}{c}{L3} & \multicolumn{1}{c|}{L4} & \multicolumn{1}{c}{L1} & \multicolumn{1}{c}{L2} & \multicolumn{1}{c}{L3} & \multicolumn{1}{c|}{L4} & \multicolumn{1}{c}{L1} & \multicolumn{1}{c}{L2} & \multicolumn{1}{c}{L3} & \multicolumn{1}{c|}{L4} & \multicolumn{1}{c}{L1} & \multicolumn{1}{c}{L2} & \multicolumn{1}{c}{L3} & \multicolumn{1}{c|}{L4} & \multicolumn{1}{c}{L1} & \multicolumn{1}{c}{L2} & \multicolumn{1}{c}{L3} & \multicolumn{1}{c}{L4} \\
    \midrule
    \multicolumn{1}{l|}{\multirow{7}[4]{*}{Holistic}} 
    & GPT-4o     & \colorcell{63.64}  &\colorcell{40.97}  &\colorcell{54.26}  &\colorcell{40.31} &\colorcell{52.78} &\colorcell{20.33} &\colorcell{40.55} & \colorcell{16.24}  & \colorcell{7.35}  &\colorcell{2.42} & \colorcell{15.07}  &\colorcell{12.04} & 0.87  &0.38  &2.31   &  1.30  & 1.02  &0.61   & 1.89  & 1.80        \\
    & Claude-Sonnet-4    & \colorcell{69.39}  &\colorcell{38.46}  &\colorcell{56.12}  &\colorcell{40.05} &\colorcell{41.92} &\colorcell{18.39} &\colorcell{26.94} & \colorcell{24.18}  & \colorcell{4.98}  &\colorcell{2.29} & \colorcell{15.81}  &\colorcell{13.51} & 1.38  &0.57  &1.43   & 1.70  &1.51  &0.45   & 1.53 & 2.10      \\
    & Gemini-2.5-Flash    & \colorcell{60.70}  &\colorcell{37.33} &\colorcell{40.89} &\colorcell{42.41}  & \colorcell{28.53} & \colorcell{16.23}  & \colorcell{31.80} &\colorcell{15.89} & \colorcell{2.24} &\colorcell{2.04} & \colorcell{6.24}  &\colorcell{1.93} &  0.87  & 0.38   & 1.04  &  0.90  & 0.67  &0.28   &1.27   & 1.50 \\
          & Deepseek-V3    & \colorcell{84.24} &\colorcell{43.14} &\colorcell{44.69}  &\colorcell{44.92}  & \colorcell{68.47}  &\colorcell{21.49} &\colorcell{27.01} &\colorcell{22.90} & \colorcell{14.38}  & \colorcell{3.16}  & \colorcell{14.31} &\colorcell{1.47} & 1.62  & 0.59 & 2.35  & 0.40  & 1.61  &1.25  & 1.80  & 0.30  \\
          & Qwen3-235B-A22B   & \colorcell{15.15} &\colorcell{19.00} & \colorcell{18.09} &\colorcell{12.36} & \colorcell{13.02} &\colorcell{5.89}  &\colorcell{7.90}  &\colorcell{11.51}  & \colorcell{9.90} &\colorcell{1.86}  &\colorcell{14.30} & \colorcell{0.74} & 1.35  & 0.54  & 1.39   &0.30  & 1.29   & 0.27  & 1.45  & 0.50 \\
          & Qwen2.5-Coder-7B    & \colorcell{52.73} &\colorcell{28.30}  &\colorcell{13.50} &\colorcell{4.82}  & \colorcell{17.02} &\colorcell{9.94} &\colorcell{1.79} &\colorcell{11.41} &\colorcell{2.56} &\colorcell{1.86} &\colorcell{1.77} &\colorcell{0.00} & 0.21  &0.38  &0.82   &0.40  & 0.47  & 0.35   & 0.41 & 0.30  \\
\cmidrule{2-22}          & Average   & \colorcell{57.64} & \colorcell{34.53} & \colorcell{37.93} & \colorcell{30.81} & \colorcell{36.96} & \colorcell{15.38} & \colorcell{22.67} & \colorcell{17.02} & \colorcell{6.90} & \colorcell{2.27} & \colorcell{11.25} & \colorcell{4.95} & \colorcell{1.05} & \colorcell{0.47} & \colorcell{1.56} & \colorcell{0.83} & \colorcell{1.10} & \colorcell{0.54} & \colorcell{1.39} & \colorcell{1.08} \\
    \midrule
    \multicolumn{1}{l|}{\multirow{7}[4]{*}{Incremental}}
    & GPT-4o    & \colorcell{66.06} &\colorcell{41.64} &\colorcell{54.52} &\colorcell{40.00}  & \colorcell{46.64} &\colorcell{19.50} & \colorcell{18.09}  & \colorcell{4.52}  & \colorcell{3.83} & \colorcell{1.86} & \colorcell{9.89}  & \colorcell{2.15} & 0.74 &  0.31 &  1.47  & 0.30  & 0.74  & 0.77  & 1.63  & 0.40   \\
    & Claude-Sonnet-4    & \colorcell{76.36}  &\colorcell{37.03}  &\colorcell{50.93}  &\colorcell{46.96} &\colorcell{60.56} &\colorcell{16.98} &\colorcell{15.29} & \colorcell{8.04}  & \colorcell{9.90}  &\colorcell{2.69} & \colorcell{9.83}  &\colorcell{2.46} & 1.24  &0.42 &1.24   & 0.50  &1.71  &0.56  & 1.58 & 0.30      \\
     & Gemini-2.5-Flash    & \colorcell{72.12} & \colorcell{38.14} &\colorcell{55.83} &\colorcell{45.76}  & \colorcell{33.07}  &\colorcell{5.00} &\colorcell{15.71} & \colorcell{3.41}  & \colorcell{6.71} &\colorcell{1.86} & \colorcell{7.76} & \colorcell{0.00} & 1.37  &0.69 & 1.83  &0.00  & 1.02  & 0.00  & 1.20  & 0.00  \\
          & Deepseek-V3    & \colorcell{69.09} &\colorcell{42.32} &\colorcell{53.08}  &\colorcell{45.65}  &\colorcell{34.21}  & \colorcell{15.34} &\colorcell{21.25}  & \colorcell{10.09}  & \colorcell{6.39}  &\colorcell{2.04}   & \colorcell{3.96}  & \colorcell{0.86} & 1.17  & 0.58 &  0.71  & 0.40  &  1.26  &0.45  &0.00 & 0.40   \\
          & Qwen3-235B-A22B   & \colorcell{26.06} &\colorcell{23.58} &\colorcell{14.68} &\colorcell{43.87}  & \colorcell{14.99} & \colorcell{8.68} &\colorcell{2.82}  & \colorcell{4.31} & \colorcell{6.71} & \colorcell{2.23}  &\colorcell{2.73} & \colorcell{1.52} & 1.05  & 0.29  & 0.45 &  0.30  & 0.82  & 0.31  & 0.40  & 0.50 \\
    & Qwen2.5-Coder-7B    & \colorcell{50.91} &\colorcell{31.54}  &\colorcell{31.98}  &\colorcell{48.06} & \colorcell{20.62}  &\colorcell{8.03} &\colorcell{4.47}  & \colorcell{5.39}  &\colorcell{6.43}  &\colorcell{1.49} & \colorcell{2.59}  & \colorcell{2.74} & 1.92  & 0.29   &0.45 & 0.30  & 1.03   & 0.50 &  0.26  & 0.60 \\
\cmidrule{2-22}          & Average   & \colorcell{60.10} & \colorcell{35.71} & \colorcell{43.50} & \colorcell{45.05} & \colorcell{35.02} & \colorcell{12.26} & \colorcell{12.94} & \colorcell{5.96} & \colorcell{6.66} & \colorcell{2.03} & \colorcell{6.13} & \colorcell{1.62} & \colorcell{1.25} & \colorcell{0.43} & \colorcell{1.03} & \colorcell{0.30} & \colorcell{1.10} & \colorcell{0.43} & \colorcell{0.85} & \colorcell{0.37} \\
    \midrule
    \multicolumn{1}{l|}{\multirow{7}[4]{*}{RAG}} 
    & GPT-4o     & \colorcell{29.09}  &  \colorcell{24.99}   &  \colorcell{34.30}  &  \colorcell{38.48}   & \colorcell{24.55} &  \colorcell{6.27}  & \colorcell{9.86}  &\colorcell{13.15}  & \colorcell{13.19} &\colorcell{2.26} & \colorcell{1.09} & \colorcell{1.43}   &  1.80  &0.34  &  0.79  & 0.70  &1.51   & 0.56  & 1.03  & 1.00 \\
    & Claude-Sonnet-4    & \colorcell{28.79}  &\colorcell{27.41}  &\colorcell{15.60}  &\colorcell{42.31} &\colorcell{26.03} &\colorcell{7.27} &\colorcell{7.31} & \colorcell{6.41}  & \colorcell{14.38}  &\colorcell{1.43} & \colorcell{5.54}  &\colorcell{5.49} & 1.67  &0.31 &0.53   & 0.40 &1.83  &0.27  & 0.64 & 0.50     \\
    & Gemini-2.5-Flash  & \colorcell{20.04}  &\colorcell{10.50} & \colorcell{24.51} & \colorcell{35.04}  & \colorcell{12.73}  & \colorcell{3.13} & \colorcell{11.60}  & \colorcell{8.93} & \colorcell{9.42}  &\colorcell{1.85}  & \colorcell{4.75} &  \colorcell{3.47} &  0.94  &  0.29  &  0.68 &  0.40 & 0.81   & 0.47  &  0.62 &  0.20  \\
          & Deepseek-V3    & \colorcell{26.67} &\colorcell{29.97}  &\colorcell{28.70}  & \colorcell{19.27}  &    \colorcell{19.34}    &  \colorcell{6.98}   &     \colorcell{25.90}    &  \colorcell{9.07}  & \colorcell{9.27}   &  \colorcell{2.51}   &  \colorcell{14.18}  &  \colorcell{8.60} & 1.03  & 0.46  & 1.15  & 0.50  &1.35   & 0.19  &  0.83  & 0.30   \\
          & Qwen3-235B-A22B   & \colorcell{29.09} & \colorcell{20.75}   & \colorcell{20.28}  & \colorcell{30.21}  & \colorcell{21.19}  & \colorcell{3.67}   & \colorcell{5.08}   & \colorcell{8.43}  & \colorcell{12.46} & \colorcell{1.79} & \colorcell{4.57} & \colorcell{2.49}   & 1.47  &  0.38 & 0.76  &  0.30 & 1.62  & 0.43  &0.71   & 0.00  \\
     & Qwen2.5-Coder-7B  & \colorcell{43.03} &\colorcell{23.45} &\colorcell{17.82}  & \colorcell{26.32}    & \colorcell{30.65}      &\colorcell{16.04}   & \colorcell{12.09}   & \colorcell{15.50}    & \colorcell{17.82}   & \colorcell{4.79} & \colorcell{2.45}  & \colorcell{4.43} & \colorcell{3.15}  & 0.75 & 0.52  &  0.40 & 0.50 & 0.74  &0.46   & 0.40 \\
\cmidrule{2-22}          & Average   & \colorcell{29.45} & \colorcell{22.85} & \colorcell{23.54} & \colorcell{31.94} & \colorcell{22.42} & \colorcell{7.23} & \colorcell{11.97} & \colorcell{10.25} & \colorcell{12.76} & \colorcell{2.44} & \colorcell{5.43} & \colorcell{4.32} & \colorcell{1.68} & \colorcell{0.42} & \colorcell{0.74} & \colorcell{0.45} & \colorcell{1.27} & \colorcell{0.44} & \colorcell{0.72} & \colorcell{0.40} \\
    \bottomrule
    \end{tabular}%
  \label{tab:ablation}%
  }
\end{table}%

\subsection{RQ4: Bad Case Analysis}

We conduct an in-depth analysis of error cases to find more useful insights into LLMs' capabilities and limitations in repo-level code generation. Specifically, we demonstrate and discuss representative error cases according to the performance of our designed five evaluation metrics.

\subsubsection{Requirement Errors}

System design defines functional objectives and constraints for the repository. When LLMs-generated implementations violate core requirements, requirement errors occur. We identify three primary error types based on severity and nature.

\textbf{\textit{Functional Misalignment}}.
LLMs outputs often do not match what users want. The generated code may include unnecessary features while missing important required functions. For example, in one real case, the requirement asked for a ``Chroma Auditor'' to check text chunks and metadata in vector databases. However, the generated code only provided basic collection management functions like creating and deleting collections and cleaning orphaned UUIDs. It completely missed the main auditing purpose. This problem may happen because long prompts contain too many technical details that hide the main system design goals.

\textbf{\textit{Core Functionality Absence}}. 
Generated code often misses important parts needed for business logic by providing unfinished core modules or placeholder code that does nothing. This problem shows a key weakness of LLMs: they cannot handle complex workflows when requirements need multiple components to work together. This leads to failures in separate modules that should connect with each other.

 \textbf{\textit{Failure to Meet Non-functional Requirements}}.  
 Generated code often does not follow the non-functional requirements like creating security problems. For example, in one real case, a user needed a modular tool to check database schema changes and stop unsafe operations. However, the LLM-generated code had problems. The SquawkLinter module used \textit{shell=True} for subprocess execution (through \textit{squawk.py}), which created a command injection security risk. Also, the generated solution did not work on different platforms and could only run on Linux and macOS systems.

\subsubsection{System Design Errors}
System design establishes structural patterns and processing workflows for the repository.  When LLM-generated implementations violate architectural principles or mishandle institutional relationships, institutional errors occur.  We identify four primary error types based on their impact scope.

\textbf{\textit{Functional Deficiencies in Modules}}. LLMs implementations omit critical functions specified in design specifications.  Essential capabilities defined in UML diagrams remain unimplemented, creating systemic gaps. This may be attributed to the difficulties LLMs face in decomposing composite functions within a system architecture into atomic operations.

\textbf{\textit{Architectural Design Flaws}}. Generated code exhibits structural inconsistencies with intended architecture patterns. This includes improper layer separation and responsibility misalignment. This discrepancy may stem from LLMs misinterpreting ``modules'' as physical files rather than logical units, resulting in a misalignment of functional responsibilities.

\textbf{\textit{Data Processing Inconsistencies}}. LLMs' implementations fail to maintain uniform data handling across system boundaries. Inconsistent format support and processing logic create integration fault lines.  For instance, Student registration supports both form-data and JSON, while teacher registration only accepts form-data.

\textbf{\textit{Type Information Discrepancies}}. Generated code exhibits semantic gaps between design specifications and implementation types. This discrepancy may arise when the UML specification marks variables as a null type, whereas the implementation defines them with concrete types.

\begin{figure}[htbp]
    \centering
    \begin{minipage}{0.48\textwidth}
        \centering
        \includegraphics[width=\textwidth]{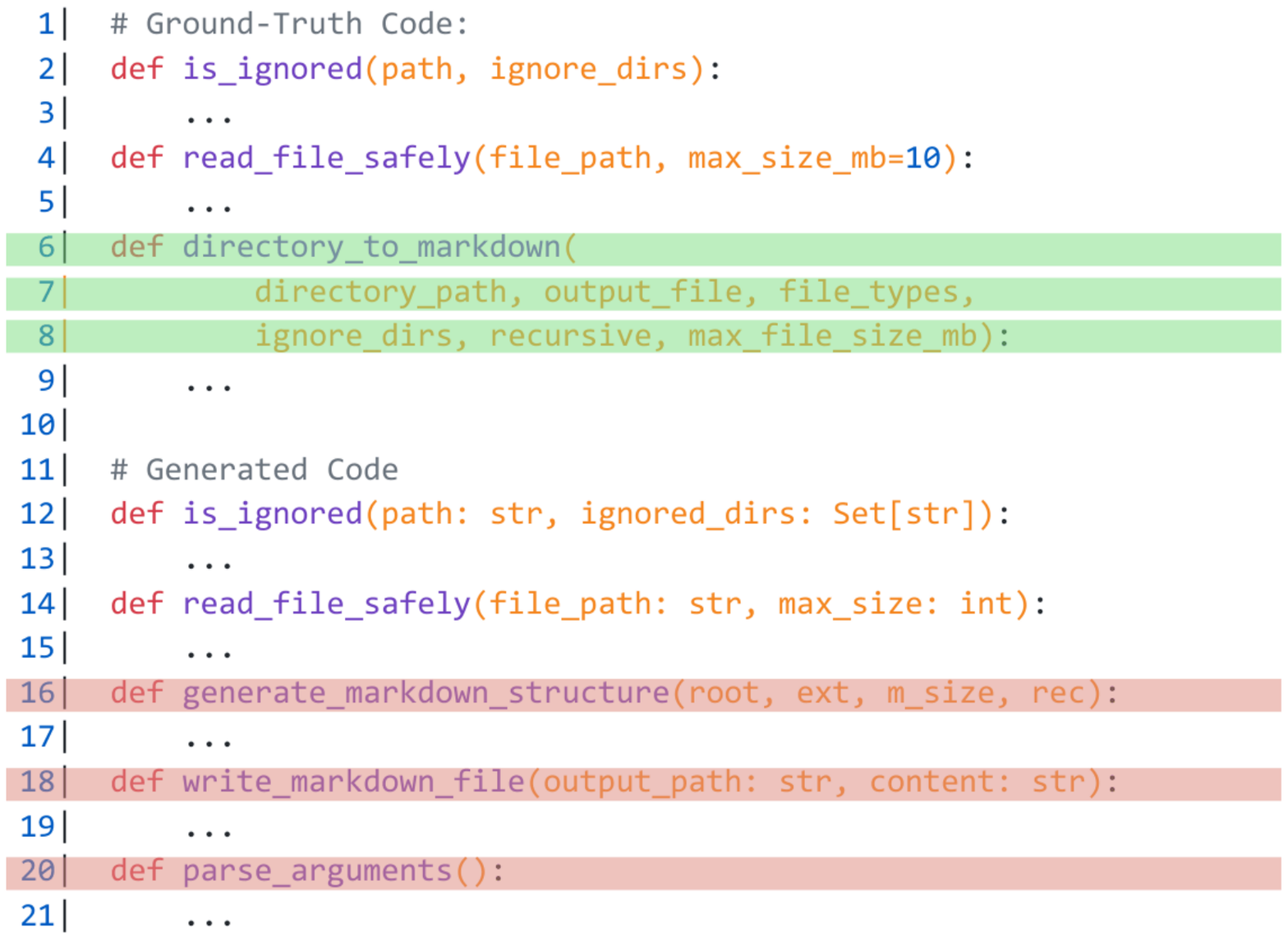}
        \caption{Module Deficiency and Module Redundancy}
        \label{Completion Errors 1}
    \end{minipage}
    \hfill
    \begin{minipage}{0.48\textwidth}
        \centering
        \includegraphics[width=\textwidth]{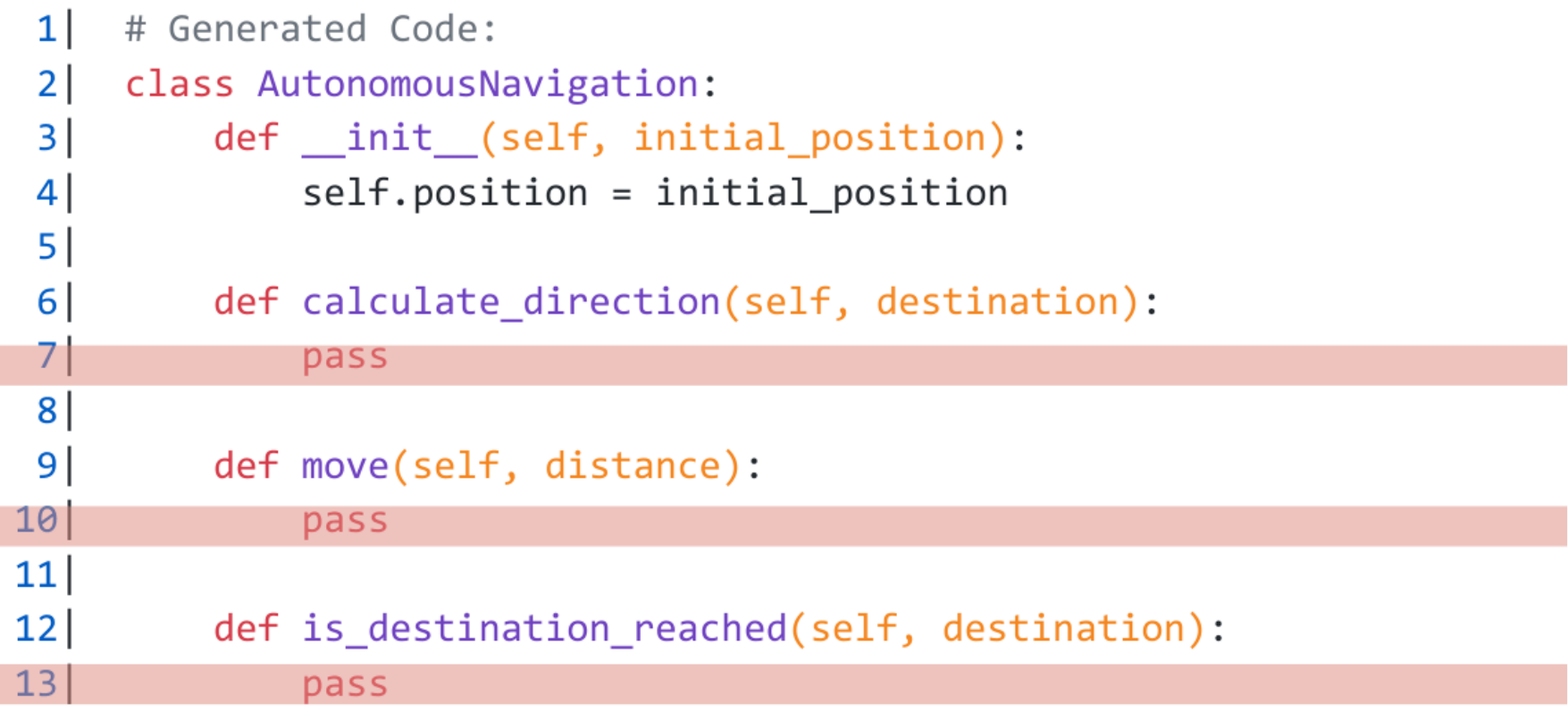}
        \caption{Failure to Generate Module Bodies}
        \label{Completion Errors 2}
    \end{minipage}
\end{figure}

\subsubsection{Completion Errors}

System design provides modules, attributes, and public exports of a repository. Ideally, LLMs should be capable of generating each module. If LLMs neglect to generate modules, a completion error occurs. There are mainly two types of typical errors. 

\textbf{\textit{Module Deficiency and Redundancy}}. LLMs usually generate redundant modules that are needless for implementing users' requirements, while lacking some modules compared to counterparts provided in system designs. As shown in Figure \ref{Completion Errors 1}, LLMs generate the needless modules ``generate$\_$markdown$\_$structure'', ``write$\_$markdown$\_$file'', and ``parse$\_$arguments''. The function ``directory$\_$to$\_$markdown'' fails to be generated. The reason might be that LLMs can not identify all modules from the system design in the long prompt.

\textbf{\textit{Failure to Generate Module Bodies}}. Although the module names and their attributes are generated correctly, LLMs only generate ``pass'' as the module's body, as shown in Figure \ref{Completion Errors 2}, resulting in incomplete functionality of the repository.

\begin{figure}[tbp]
    \centering
    \begin{minipage}{0.48\textwidth}
        \centering
        \includegraphics[width=\textwidth]{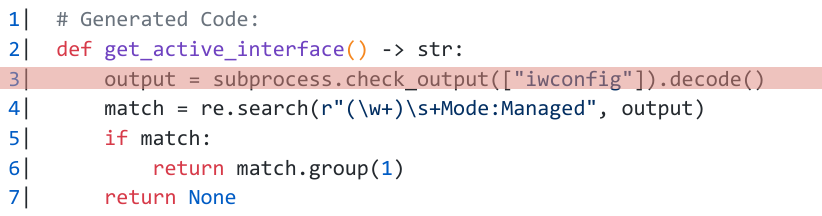}
        \caption{Incorrect Attribute Usage}
        \label{F: Incorrect Attribute Usage}
    \end{minipage}
    \hfill
    \begin{minipage}{0.48\textwidth}
        \centering
        \includegraphics[width=\textwidth]{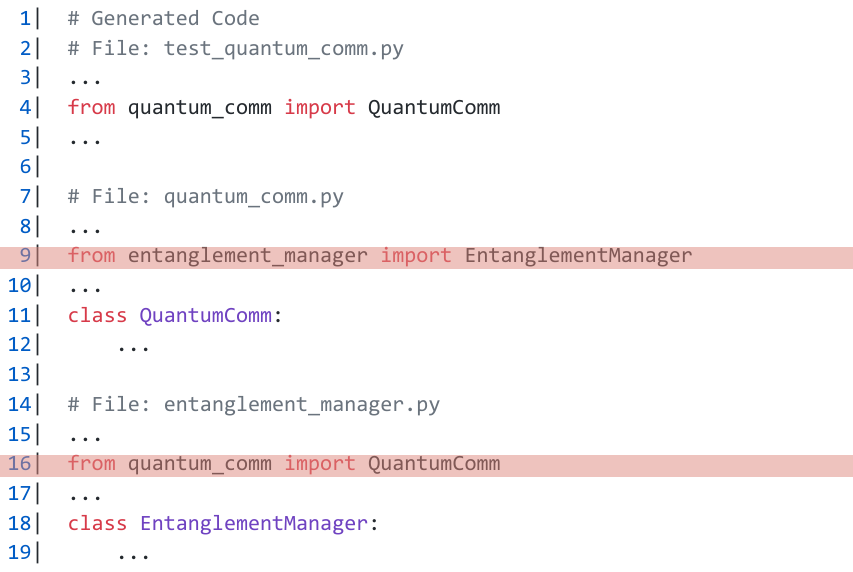}
        \caption{Incorrect Invoking and Inheritance}
        \label{F: Incorrect Invoking and Inheritance}
    \end{minipage}
\end{figure}

\subsubsection{Execution Errors}

Execution errors focus on assessing the syntax and format of generated repositories. 
The statistics show that execution errors are mainly related to incorrect attribute usage and incorrect invocation or inheritance among code snippets. 

\textbf{\textit{Incorrect Attribute Usage}}.   
The attributes of code elements are crucial for the correct execution of repositories. However, the verified LLMs often make mistakes in identifying attributes of code. As shown in Figure \ref{F: Incorrect Attribute Usage},  the return type of ``check$\_$output'' method is a string, while the string can not be processed by ``decode()''. This mistake results in a syntax error, causing the repository to fail during execution.

\textbf{\textit{Incorrect Invoking and Inheritance}}. Another common phenomenon is that LLMs incorrectly invoke and inherit predefined code snippets in a repository. 
In Figure \ref{F: Incorrect Invoking and Inheritance}, the ``entanglement$\_$manager'' module invokes the ``quantum$\_$comm'' module and vice versa. The cyclic invoking between modules leads to the failure to start the project. This reflects that LLMs' ability to understand system design and analyze relationships among modules needs to be further enhanced, which provides an insightful direction for LLM optimization.

\subsubsection{Pass Errors}

\begin{figure}[htbp]
    \centering
    \begin{minipage}{0.48\textwidth}
        \centering
        \includegraphics[width=0.85\textwidth]{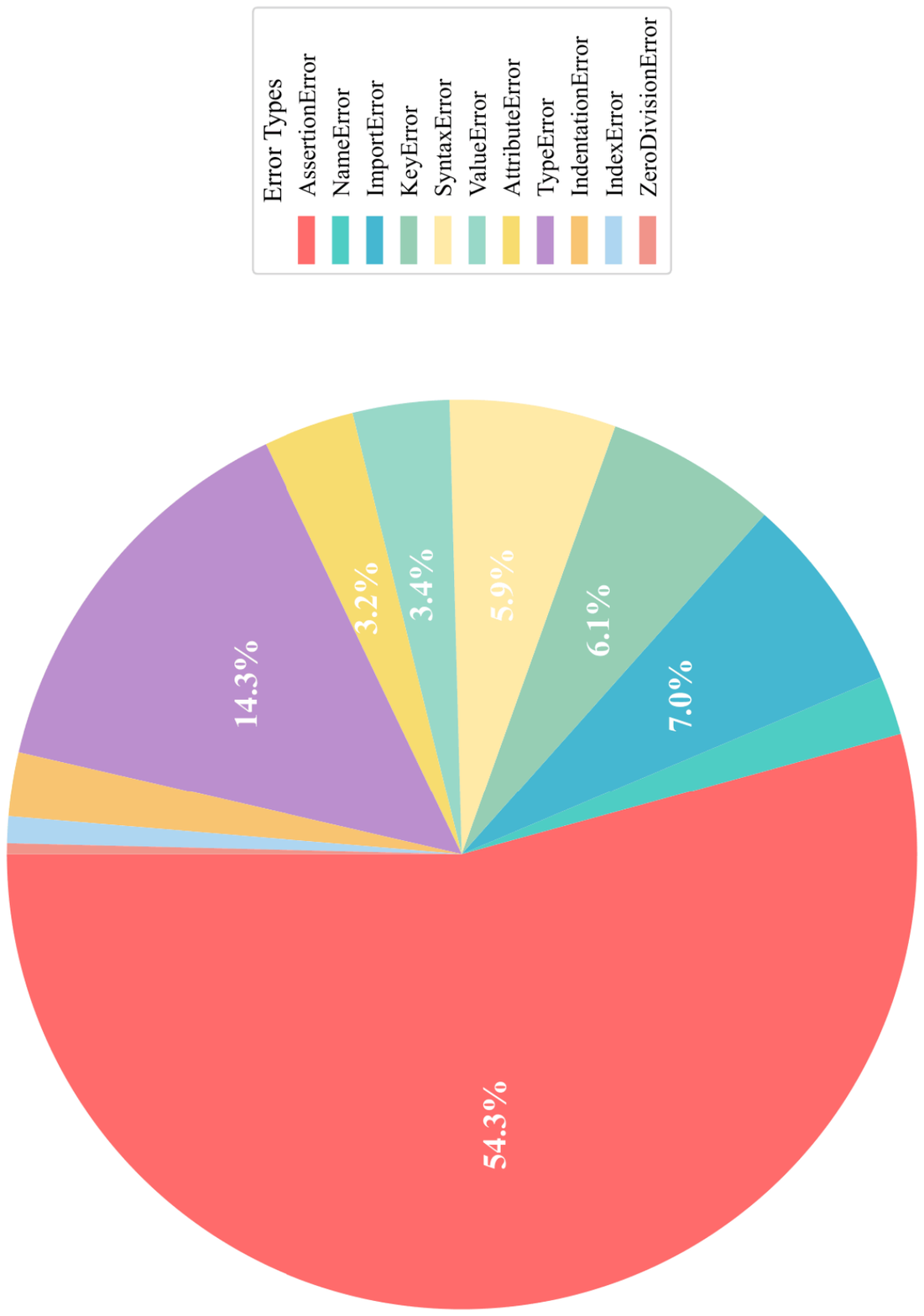}
        \caption{Error Type Distribution of Generated RealBench on test pass rate.}
        \label{error_types}
    \end{minipage}
    \hfill
    \begin{minipage}{0.48\textwidth}
        \centering
        \includegraphics[width=\textwidth]{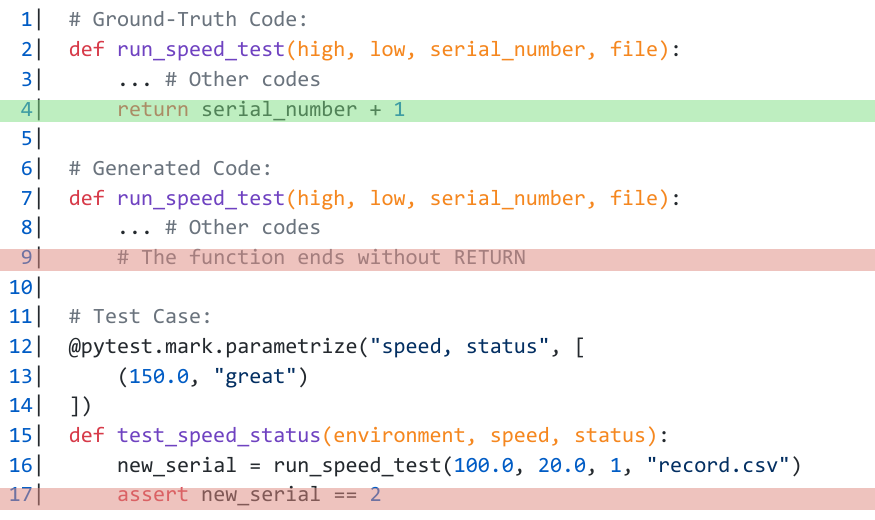}
        \caption{Incorrect Logic}
        \label{F: Incorrect Logic}
    \end{minipage}
\end{figure}

This error indicates that the generated code fails to pass test cases, which is mainly caused by incorrect functionalities or syntax. We analyze the error logs and provide the type distribution of errors as shown in Figure~\ref{error_types}. There are mainly 11 error types in the generated repositories. Different colors represent the different error types.

Among these 11 errors, 2 of them may prevent the code from running, i.e., \textit{SyntaxError} and \textit{IndentationError}, which contribute a total of 9.1\% of total errors. For the remaining errors that were raised at runtime, \textit{AssertionError} is the most dominating one, which contributes to 54.3\% of total errors, while \textit{TypeError} serves as the second largest category, which contributed 14.3\% of errors.
Then we manually analyzed the typical errors made by LLMs that caused these runtime errors, which mainly consist of 2 types.

\textbf{\textit{Incorrect Logic}}. The generated code contains flawed implementation logic that produces wrong results or behavior. For the example in Figure~\ref{F: Incorrect Logic}, the generated function ends without a return statement, causing it to return None instead of the expected value. When the test case expects a value of 2, it fails because the function returns None rather than the computed value, which caused an \textit{AssertionError}.

\textbf{\textit{Incorrect Arguments}}. 
The generated code uses wrong parameter names, types, or default values in function signatures. This type of error occurs when the code generator fails to match the expected function interface, such as renaming parameters, changing their types, or omitting default values. These signature mismatches break the function's calling convention and result in \textit{TypeError} exceptions when the function is invoked with arguments that no longer align with the generated parameters.

\vspace{1mm}
\begin{custommdframed}
\textbf{Finding 6:} \textit{AssertionError} and \textit{TypeError}  are the Top-2 dominating errors, contributing to 68.6\% (= 54.3\% + 14.3\%) of total errors. 
\end{custommdframed}
\vspace{0mm}

%% file: sec5-analysis.tex
\section{Threats to Validity}

This paper has several main threats to validity. 
First is the threats in dataset construction. The quality and detail level of natural language descriptions could affect LLMs' code generation performance, and selecting representative topics may introduce bias. To alleviate this threat, we carefully selected representative topics and manually verified all requirements, architectures, and test cases in our samples to ensure dataset quality.
Second is on data leakage. LLMs may have seen similar code during training, which could lead to inflated evaluation scores. To prevent this issue, we collected all projects in our dataset from GitHub after 2024-12, which is earlier than the release time of evaluated models to mitigate data leakage problems.
Third is on the impact of model prompting. Choosing an optimal prompt is challenging, and different prompts could yield different performance results. To mitigate this threat, we follow common practices for prompting LLMs to design our prompt templates. We also report results using greedy decoding, which is deterministic and reduces randomness in model responses.

%% file: sec6-relatedwork.tex
\section{Related Work}

In this section, we discuss the related works on code generation and the benchmarks.

\noindent\textbf{Code Generation.}
Recent work on code generation has focused on using LLM to generate code. Several studies~\cite{jiang2024completionsurvey,husein2025completionreview,izadi2024completionevaluation,nguyen2019function} evaluate how well LLMs perform this task. Tools like CodeFill~\cite{izadi2022codefill} and aiXcoder~\cite{jiang2024aixcoder} use different techniques to predict code structure and improve generation speed. 
Another research area uses RAG frameworks~\cite{lewis2020rag} to generate code at the repository level. These methods find relevant code from other files in the same repository to help with generation. Several approaches including RepoCoder, GraphCoder, and DraCo~\cite{zhang2023repocoder,liu2024graphcoder, cheng2024draco} try different ways to find relevant code. They use static analysis, control flow graphs, and similarity matching. LongCoder~\cite{guo2023longcoder} takes a different approach by making the model handle longer code inputs.

\noindent\textbf{Code Generation Benchmarks.}
To better understand the capability of code generation approaches, researchers have created many benchmarks, which can be categorized to function-level, class-level and repository-level.

\textit{Function-level Benchmarks.} Early code generation benchmarks focus on generating single functions or code snippets. HumanEval~\cite{chen2021evaluating} is one of the most widely used benchmarks in this category. It contains 164 Python programming problems that require generating functions based on natural language descriptions. Similarly, MBPP~\cite{austin2021program} provides another function-level benchmark with basic Python programming problems. APPS~\cite{hendrycks2021measuring} extends the difficulty by including competition-style programming problems that require more complex algorithmic thinking.

\textit{Class-level Benchmarks.} As LLMs became more capable, researchers introduced benchmarks that require generating entire classes or more complex code structures. ClassEval~\cite{du2023classeval} evaluates LLMs on class-level code generation with 100 human-crafted self-contained Python classes. This benchmark tests the ability to generate object-oriented code with multiple methods and proper class structure.  Benchmarks like Concode~\cite{iyer2018mapping} and CoderEval~\cite{yu2024codereval} focus on non-standalone programs that depend on external libraries or frameworks. 

\textit{Repository-level Benchmarks.} The most recent and challenging benchmarks evaluate LLMs on repository-scale code generation tasks. These benchmarks require understanding complex codebases and generating code that fits within larger software projects. JavaBench~\cite{cao2024javabench} asks LLMs to generate entire repositories based on natural language requirements. EvoCodeBench~\cite{li2024evocodebench} also falls into this category by requiring the generation of functions or classes within a given repository context.
However, existing repo-level code generation benchmarks only provide natural language requirements for each task. They ignore the common industry practice where developers usually implement code based on both design documents and requirements rather than raw requirements.

\section{Conclusion}

In this paper, we propose RealBench, a repository-level code generation benchmark aligned with real-world software development practices. It provides both natural language requirements and UML diagrams as system design, matching how developers receive specifications in industry settings. RealBench consists of 61 repositories across 20 domains, each with comprehensive test suites achieving 79.76\% line coverage. To assess generated repositories, we design a systematic evaluation framework with two granularities and five metrics. Our experiments on six advanced LLMs using three generation strategies show several meaningful insights.

\section*{Acknowledgment}
This research is supported by the National Natural Science Foundation of China under Grant No. 92582203, 62192731, 62192730, 62192733, 62072007, the National Key R\&D Program under Grant No. 2023YFB4503801, the Beijing Major Science and Technology Project under Contract No. Z251100008425005.